\documentclass[journal]{IEEEtran}
\usepackage[english]{babel}
\usepackage[square, comma, sort&compress, numbers]{natbib}
\usepackage{graphicx} 
\usepackage{ntheorem,amsmath,bm}
\usepackage{amssymb,amsfonts}
\usepackage{graphicx}
\usepackage{textcomp}
\usepackage{xcolor}
\usepackage{orcidlink}
\usepackage{float}
\usepackage{algorithm}
\usepackage{algorithmic}
\usepackage{subcaption,caption}

\IEEEoverridecommandlockouts
\newtheorem{thm}{Proposition}
\newtheorem{remark}{Remark}
\theoremheaderfont{\it}\theorembodyfont{\upshape} \theoremstyle{nonumberplain}
\theoremseparator{.} \theoremsymbol{\rule{1ex}{1ex}} \newtheorem{proof}{Proof}

\def \argmax {{ \rm argmax}}

\begin{document}
\title{Beampattern Synthesis for Discrete Phase RIS in Communication and Sensing Systems}
\author{\IEEEauthorblockN{Xiao Cai \orcidlink{0009-0002-2291-5757}, Hei Victor Cheng \orcidlink{0000-0001-8432-3779} and  Daniel E. Lucani \orcidlink{0000-0001-5325-8863}}\\
\IEEEauthorblockA{Department of Electrical and Computer Engineering, Aarhus University, Denmark\\ 
Email: xiao.cai@ece.au.dk, hvc@ece.au.dk, daniel.lucani@ece.au.dk}

\thanks{Hei Victor Cheng is the corresponding author. The materials in this paper have been presented in part at the 2024 IEEE 25th International Workshop on Signal Processing Advances in Wireless Communications (SPAWC) \cite{cai2024array}. This work is supported in part by the Aarhus Universitets Forskningsfond project number AUFF 39001 and the NordForsk Nordic University Cooperation on Edge Intelligence (Grant No. 168043)
.}
}

\maketitle

\begin{abstract}
Extensive research on Reconfigurable Intelligent Surfaces (RIS) has primarily focused on optimizing reflective coefficients for passive beamforming in specific target directions. This optimization typically assumes prior knowledge of the target direction, which is unavailable before the target is detected. To enhance direction estimation, it is critical to develop array pattern synthesis techniques that yield a wider beam by maximizing the received power over the entire target area. Although this challenge has been addressed with active antennas, RIS systems pose a unique challenge due to their inherent phase constraints, which can be continuous or discrete. 

This work addresses this challenge through a novel array pattern synthesis method tailored for discrete phase constraints in RIS. We introduce a penalty method that pushes these constraints to the boundary of the convex hull. Then, the Minorization-Maximization (MM) method is utilized to reformulate the problem into a convex one. Our numerical results show that our algorithm can generate a wide beam pattern comparable to that achievable with per-power constraints, with both the amplitudes and phases being adjustable. We compare our method with a traditional beam sweeping technique, showing a) several orders of magnitude reduction of the MSE of Angle of Arrival (AOA) at low to medium Signal-to-Noise Ratio (SNR)s; and b) $8$~dB SNR reduction to achieve a high probability of detection.  
\end{abstract}


\section{Introduction}

Reconfigurable Intelligent Surfaces (RIS) \cite{liu2021reconfigurable,bjornsonReconfigurableIntelligentSurfaces2022} represent an emerging technology that dynamically controls the reflection of signals using numerous antenna elements that can shape the beam to enhance various performance metrics of communication systems \cite{wuIntelligentReflectingSurface2019,bai2020latency,pan2020multicell}. A common assumption is that RIS acts as a passive antenna array, which involves adjusting the phase of each RIS element \cite{lebret1997antenna}. In practice, hardware limitations often impose discrete phase constraints, meaning that phase shifts applied to RIS elements can only take on certain discrete values \cite{PhaseonlySynthesisAntenna,smithComparisonMethodsRandomizing1983,ismailArrayPatternSynthesis2010}. This constraint presents challenges for the optimization method as it is NP-hard in general, thus necessitating specialized methods to achieve optimal performance \cite{wuBeamformingOptimizationWireless2020,diHybridBeamformingReconfigurable2020,zhangConfiguringIntelligentReflecting2022,renLinearTimeAlgorithm2023}. 

\textcolor{black}{
From both communication and sensing perspective, wide beams are of great importance.} In the case of sensing, target detection requires designing the RIS be a wide beam since high-gain beams that depend on directional alignment may miss the target. A wide beam provides uniform illumination over a broad angular region, increasing the probability of detection \cite{orchard1985optimising}.
\textcolor{black}{
In the case of communication, another application arises in the initial access and synchronization phase, where many users are dispersed across wide sectors at unknown locations. }The traditional approach employs beam sweeping \cite{giordaniComparativeAnalysisInitial2016,giordaniInitialAccess5G2016,giordaniTutorialBeamManagement2018}, where the RIS or Base Station (BS) sequentially employs narrow beams to scan all potential angular regions. While effective, beam sweeping requires multiple pilot transmissions, which increases latency and control overhead. In contrast, a wide beam can deliver a cell-specific or sector-specific reference signal with a single pilot, achieving simultaneous coverage and faster access.

\subsection{Related Work}
Studies have investigated the role of RIS in communication, highlighting its potential to enhance performance in various scenarios. For instance, \cite{wuIntelligentReflectingSurface2019} introduced a joint active and passive beamforming technique, demonstrating that an RIS-aided MIMO system can achieve the same rate performance as a massive MIMO system without RIS. In another application, \cite{cheng2023degree} utilized RIS to transmit signals, thereby improving the degrees of freedom (DoF) of the overall channel. Furthermore, RISs have been deployed to strengthen the signal power for cell-edge users and to mitigate inter-cell interference \cite{pan2020multicell}.

Beyond communications, RIS is gaining significant interest for sensing applications, particularly within the Joint Communication and Sensing (JCAS) framework. This trend leverages communication systems for sensing tasks like indoor localization \cite{zafari2019survey}, Wi-Fi-based activity recognition \cite{li2017indotrack}, and automotive radar \cite{ma2020joint}. \textcolor{black}{
Recent works have explored the use of RIS to enhance sensing performance, particularly in scenarios where LoS paths are weak or blocked. By redirecting and strengthening reflected signals, RIS has the potential to support more reliable detection \cite{aubry2021reconfigurable,buzzi2022foundations} and localization \cite{kim2023ris, kemal2024ris}.}

For example, in target detection, \cite{buzzi2022foundations} develops a signal model for monostatic and bistatic radar configurations that incorporates up to two RIS where they optimize RIS phase shifts to maximize detection probability. 
In angle estimation, \cite{kim2023ris} presents an RIS-enabled SLAM framework, which includes an adaptive RIS design and the derivation of theoretical bounds for channel parameter estimation. Subsequently, \cite{kemal2024ris} extended this approach to the detection of moving targets.

\textcolor{black}{
These schemes mentioned above all optimize the RIS beam pattern toward a specific direction of interest, resulting in a narrow beam, which becomes ineffective in scenarios where the target location is unknown or dynamic.}
This limitation has motivated research into alternative strategies. 
One such direction is blind beamforming: \cite{laiBlindBeamformingIntelligent2023} uses random weights to sense the environment. A more direct approach is to design a wide beam that provides uniform coverage over a large area. Such wide beams are crucial for robust channel estimation \cite{zhang2021overview,chenjie2023channelesti,cai2023direct} and for sensing systems \cite{huMassiveMIMOPotential2018,wymeerschRadioLocalizationMapping2020}. 

\textcolor{black}{
Recent work has begun to address the design of wide beams for RIS.} For instance, \cite{luanPhaseDesignNearField2022} tackles the beam-split problem in near-field wideband systems by dividing the coverage area into grids, thus creating a wide and flat gain beam. In \cite{al-tousCoverageAreaOptimized2024}, the authors define an ideal wide beam shape using discrete prolate spheroidal sequences and then formulate an optimization problem to minimize the mean squared error between the RIS-generated pattern and this ideal shape. From a different perspective, \cite{ramezaniBroadBeamReflectionDualPolarized2024} leverages the properties of Golay complementary sequences to synthesize a wide beam.
\textcolor{black}{ 
While effective in their respective scenarios, these methods are typically tailored to fixed beamwidths determined by the chosen template or coverage grid, and lack the flexibility to produce beams with variable widths as needed.}

\textcolor{black}{ Parallel to wide-beam synthesis, a separate line of work focuses on conventional narrow-beam formation under some hardware constraints }such as \cite{wang2012design} and \cite{caoConstantModulusShaped2017}. Yet, these methods assume continuously adjustable phase values, which are incompatible with practical RIS hardware that supports only discrete phase shifts \cite{youChannelEstimationPassive2020}. Although recent advances have provided optimal discrete-phase beamformers based on geometric interpretations \cite{zhangConfiguringIntelligentReflecting2022,renLinearTimeAlgorithm2023}, \textcolor{black}{they are limited to synthesizing highly focused, single-direction beams.}

\textcolor{black}{Thus, existing RIS beam design methods either (i) support only a fixed beamwidth (wide or narrow), or (ii) ignore the discrete phase constraint in practical RIS. As a result, no prior work addresses the general problem of synthesizing beams with arbitrary beamwidth under discrete constraints.}

\vspace{-0.37cm}
\subsection{Main Contribution}
In this paper, the primary problem we aim to solve is the synthesis of beams with a given beamwidth in RIS under discrete-phase constraints. Traditional methods, which assume continuous phases, are not directly applicable. There remains a significant gap in achieving wide beams. This gap highlights the need for innovative solutions that can efficiently handle the discrete phase constraint while providing broad coverage.
Our main contributions are summarized as follows:
\begin{itemize}
  \item We formulate and address the problem of synthesizing wide beams for RIS under discrete phase constraints.
  \item \textcolor{black}{Our algorithm supports flexible beamwidth control while satisfying discrete phase resolution limits, offering a more general design framework compared to existing wide-beam or discrete-phase-only methods.}
  \item We provide comprehensive simulation results to verify the effectiveness and robustness of our proposed algorithm under various system settings.
  \item We demonstrate that the proposed wide beam outperforms the traditional beam sweeping method in downstream tasks such as angle estimation and target detection.
\end{itemize}

\subsection{Organization \& Notations}

The remainder of the paper is organized as follows. Section \ref{sec:2} describes the system model used in our study, including assumptions and parameters. Section \ref{sec3} outlines the problem of wide beam synthesis and the constraints involved. Then, we present our novel optimization algorithm for solving the problem under discrete phase constraints in Section \ref{sec4}. The simulation results of our algorithm are presented in Section \ref{secbmsim}. Section \ref{sec5} showcases practical applications of our method, verifying its importance and effectiveness in object detection and AOA estimation. Finally, we conclude with a summary of our findings and future research directions in Section \ref{sec6}.

\textit{Notations:} We use the following notation throughout the paper: The scalar, vector, and matrix are lowercase, bold lowercase, and bold uppercase, i.e., $a$, $\bm{a}$, and $\bm{A}$, respectively.
The notation $\mathcal{CN}(u, \sigma^2)$ is used for a complex Gaussian random variable with mean $u$ and variance $\sigma^2$. 
$\operatorname{diag}(\cdot)$ forms a vector into a diagonal matrix, and $\odot$, $\otimes$ represent the element-wise and Kronecker product of matrices.
The operator $\lVert \cdot \rVert_2$ represents the $\ell^2$ norm of a vector.
Additionally, $\operatorname{Conv}(\cdot)$ denotes the convex hull, and $\mathcal{P}$ 
indicates the projection operator. 
The notation $(\cdot)^T$, $(\cdot)^H$, and $(\cdot)^*$ stand for transpose, 
Hermitian transpose and conjugate of a matrix/vector, respectively.

\section{system model}
\label{sec:2}

Consider a system that includes a \textcolor{black}{target (TAR)}, a RIS, and a base station (BS). The TAR with an \textit{unknown} position is sending pilot signals $s$, reflected by the RIS with phase configurations $\bm{w} = [w_1, \cdots, w_i, \cdots,w_{N}]\in \mathbb{C}^{N \times 1}$ where $N$ is the number of elements and $w_i $ is the reflecting coefficient of the $i$-th element.
 $w_i $ is chose from a discrete set,  $w_i \in \mathcal{S}$:
\begin{equation}
    \mathcal{S} = \left\{\exp \left(  j \left(\frac{2\pi}{L}l + \frac{\pi}{L}\right)\right) , \quad l = 0,\cdots, L-1 \right\}
    \label{eq:dpc}
\end{equation} 
where $\mathcal{S}$is the set with $L$ discrete quantization levels.

\textcolor{black}{ Let $\theta$ denote the angle of arrival (AoA) at the RIS from the BS, measured with respect to the RIS array broadside. Similarly, let $\phi$ denote the angle of departure (AoD) from the RIS toward the target.} We assume that RIS and BS are calibrated \cite{cai2023direct}; thus $\phi$ is known, and $\theta$ is located in the region of interest (ROI) $[\theta_{min},\theta_{max}]$. 
We also assume that there is no Line-of-sight (LoS) path between the BS and the TAR, so the signal propagates through 2 jumps: TAR-RIS-BS, and denote the channel TAR-RIS, RIS-BS as $\bm{g}$ and $\bm{h}$. The TAR sends signal $s$ with unit power, and we write the received signal as follows:
\begin{equation}
\begin{aligned}
    &y = \bm H(\bm{w},\theta) s +n \\
    &\bm H(\bm{w},\theta) =  \bm h^H \text{diag} \left( \bm w \right) \boldsymbol g,
\end{aligned} 
\end{equation}
where $n\sim \mathcal{CN}(0, \sigma^2_n)$ is additive Gaussian noise.

\subsubsection{Uniform Linear Array (ULA)}
Let us start by considering RIS as a  Uniform Linear Array (ULA), the channel from TAR to RIS is characterized as $\boldsymbol{g} = \alpha \boldsymbol{a}(\theta) \in \mathbb{C}^{N \times 1}$, where $\alpha$ is the complex gain, and $\boldsymbol{a}(\theta)$ is the steering vector of the ULA, given by:
 \begin{equation}
  \label{eq:steeringvector}
 	\boldsymbol{a}(\theta)=\frac{1}{\sqrt{N}}\left[ 1,e^{j\frac{2\pi d}{\lambda} \sin(\theta)} , \cdots, e^{j(N-1)\frac{2\pi d}{\lambda} \sin(\theta)} \right]^{T}.
 \end{equation}
  where $d$ is the distance between adjacent elements and $\lambda$ is the wavelength of the carrier frequency.
 Similarly, we also denote the RIS-TAR channel as $\boldsymbol{h} = \beta \boldsymbol{a}(\phi) \in \mathbb{C}^{N \times 1}$ where $\beta$ is the relating complex gain.
In this case, our system can be modeled as:
\begin{equation}
    \label{eq:sysmodel}
        y =\alpha \beta \boldsymbol{a}^{H}(\phi) \text{diag} \left( \bm w \right) \boldsymbol{a}(\theta) s  = \bm h^H  \text{diag} \left( \bm w \right) \bm g s + n ,
  \end{equation}

\begin{figure}[tb]
    \centering
    \includegraphics[width=0.6\linewidth]{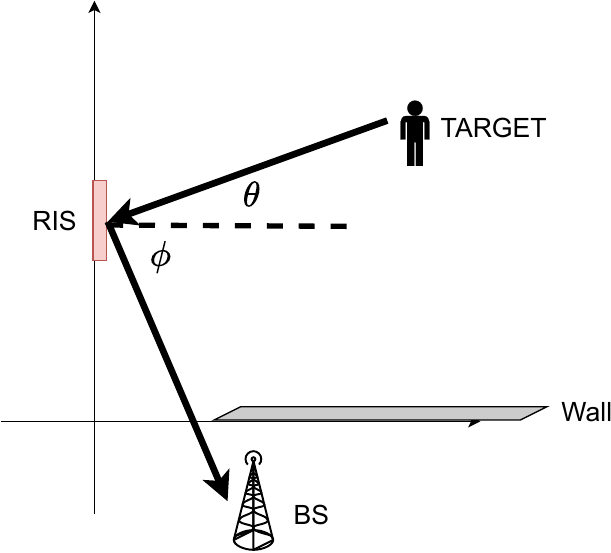}
    \caption{System setting}
    \label{fig:system}
    \vspace{-0.5cm}
\end{figure}
 
\subsubsection{Uniform Planar Array (UPA)}  In practical scenarios, the RIS is typically implemented as a planar array, which necessitates the use of the Uniform Planar Array (UPA) model for a more accurate representation.
  Let $\theta_{\rm el}$, and $\theta_{ \rm az}$ denote the elevation and azimuth angles, respectively. The array response of a UPA with $N = P \times Q$ antennas is:
  \begin{equation}
    \label{eq:UPA}
    \boldsymbol{a}(\theta_{\rm el}, \theta_{ \rm az}) = \boldsymbol{a}_y(\theta_{\rm el}, \theta_{ \rm az}) \otimes \boldsymbol{a}_z(\theta_{\rm az}),
  \end{equation}
  \begin{equation}
    \begin{aligned}
    \bm{a}_y (\psi) &\triangleq \frac{1}{\sqrt{Q}}
    \left[
    1, e^{j\frac{2\pi d}{\lambda}\psi}, \ldots, e^{j\frac{2\pi d}{\lambda}(Q-1)\psi}
    \right]^T ,\\
    \bm{a}_z (\theta_{\rm el}) &\triangleq \frac{1}{\sqrt{P}}
    \left[
    1, e^{j\frac{2\pi d}{\lambda}\sin\theta_{\rm el}}, \ldots, e^{j\frac{2\pi d}{\lambda}(P-1)\sin\theta_{\rm el}}
    \right]^T,\\
    \psi & \triangleq \sin\theta_{ \rm az}\cos\theta_{\rm el}.\\
    \end{aligned}
    \end{equation} 
Similarly, we define the system model as:
\begin{equation}
    \label{eq:sysmodel_upa}
        y = \bm h^H  \text{diag} \left( \bm w \right) \bm g s + n ,
  \end{equation}
\textcolor{black}{where  $\boldsymbol{g} = \alpha \boldsymbol{a}(\theta_{\rm el}, \theta_{\rm az}) \in \mathbb{C}^{N \times 1}$, $\boldsymbol{h} = \beta \boldsymbol{a}(\phi_{\rm el}, \phi_{\rm az}) \in \mathbb{C}^{N \times 1}$, and $\alpha, \beta$ is the complex gain.} 

 It is important to note that, although the channels differ between the ULA and UPA settings, the resulting system model expression remains generally applicable to both configurations. With this in mind, we expand the matrix multiplication and reorganize terms to obtain a unified representation:
\begin{equation}
    \label{eq:sysmodel1}
    y = \sum_{i=1}^{N} w_i g_i^{*} h_i s + n = \left(\boldsymbol{g} \odot \boldsymbol{h}^* \right)^H \boldsymbol{w} s + n = \boldsymbol{\overline{h}}^H \boldsymbol{w}s + n.
    \end{equation}
where, for convenience in subsequent formulations, we define $\left(\boldsymbol{g} \odot \boldsymbol{h}^* \right)$ as $\boldsymbol{\overline{h}}$.
Finally,  with the system model, the SNR of the received signal can be calculated as:
  \begin{equation}
    \label{eq:power}   
      \text{SNR} = \frac{1}{\sigma_n^2}  \left| \boldsymbol{\overline{h}}^H \boldsymbol{w} s\right|^2.
  \end{equation}

\section{Problem formulation}
\label{sec3}

In this section, we develop the formulation of our design problem by progressively increasing the system complexity. We begin with the simplest scenario, where the RIS is modeled as a ULA, and then extend the formulation to a UPA. Finally, we consider an extension to multiple time slots.
\paragraph{ULA-based Problem formulation}
Assume the target is located in the region of interest: $\theta \in [\theta_{\rm min}, \theta_{\rm max}]$. \textcolor{black}{ Our goal is to maximize the minimum SNR in the region of interest. This max–min criterion ensures that the worst-case performance is optimized, thereby enforcing a uniform angular power coverage.}

In the ULA setting, using the system model defined in \eqref{eq:sysmodel}, the problem of optimizing the RIS phase coefficients can be formulated as maximizing the minimum received signal power across ROI. Mathematically, the problem is stated as:
\begin{equation}
  \label{eq:opt01}
  \begin{aligned}
    \max  & \quad \min \quad  \left| \boldsymbol{\overline{h(\theta)}}^H \boldsymbol{w} \right|^2.\\
    s.t. \quad & w_i \in \mathcal{S},\  \forall i = 1, \ldots, N, 
    \quad \theta \in [\theta_{\rm min}, \theta_{\rm max}].  
  \end{aligned}
\end{equation}

\textcolor{black}{It is important to note that \eqref{eq:opt01} is a highly challenging optimization problem for several reasons. First, since the constraint must hold for every $\theta$ in the continuous interval $[\theta_{\rm min},\theta_{\rm max}]$, the problem is a semi-infinite program. Second, the objective $\min_{\theta}\left| \overline{\boldsymbol h}(\theta)^H \boldsymbol w \right|^2$ is non-convex. Third, the discrete phase constraint makes the problem combinatorial and thus NP-hard. Therefore, obtaining an exact solution to \eqref{eq:opt01} is computationally intractable, which motivates the development of efficient algorithms, which we present in the next section.}

\paragraph{UPA-based Problem formulation}
While the ULA model provides useful insights and analytical tractability, practical RIS implementations are often realized as planar arrays. In the UPA setting, the channel from the target to the RIS becomes a function of both the elevation and azimuth angles, i.e., $\theta_{\rm el}$ and $\theta_{\rm az}$. 
With the ROI now defined as $\theta_{\rm el} \in [\theta_{\rm el,min}, \theta_{\rm el,max}]$, $\theta_{\rm az} \in [\theta_{\rm az,min}, \theta_{\rm az,max}]$, the corresponding problem formulation becomes:
\begin{equation}
  \label{eq:opt09}
  \begin{aligned}
    \max \quad & \min \quad  \left| \boldsymbol{\overline{h(\theta_{\rm el}, \theta_{ \rm az})}}^H \boldsymbol{w} \right|^2\\
    s.t. \quad & w_i \in \mathcal{S},\  \forall i = 1, \ldots, N. \\
    \quad & \theta_{\rm el} \in [\theta_{\rm el,min}, \theta_{\rm el,max}], \theta_{\rm az} \in [\theta_{\rm az,min}, \theta_{\rm az,max}].
  \end{aligned}
\end{equation}
It is worth noting that the channels $\boldsymbol{g}$ and $\boldsymbol{h}$ in this case are a two-dimensional extension of the ULA concept, so is their combination $\boldsymbol{\overline{h}}=\boldsymbol{g} \odot \boldsymbol{h} ^*$. \textcolor{black}{ Although the UPA introduces two angular dimensions, its array response retains a separable structure via Kronecker products. As a result, the form of the objective and constraints in \eqref{eq:opt09} remains analogous to the ULA case, }allowing the proposed optimization framework to be directly extended. For clarity, we focus on the ULA formulation in the following discussions.


\paragraph{Extension to multiple time slots}
In practical applications such as target detection or angle estimation, transmitting multiple pilot signals can be advantageous, as it increases the overall signal energy and improves system performance. 
Suppose that the TAR sends pilot signals for $T$ time slots, $\bm{s} = [s_1, s_2, \cdots , s_T]$.
At each time slot $t$, the RIS employs a potentially different phase configuration $\boldsymbol{w}_t$, leading to the received signal:
\begin{equation}
\label{eq:sysmodeltotal}
y_t = \boldsymbol{\overline{h}(\theta)}^H \boldsymbol{w}_t s_t + n.
\end{equation}

For convenience, we define a concatenated channel:
\begin{equation}
\label{eq:e1}
\boldsymbol{\overline{f}(\theta)}^H  = \underbrace{ \left[\boldsymbol{\overline{h}(\theta)}^H, \cdots, \boldsymbol{\overline{h}(\theta)}^H \right]}_{T} \in \mathbb{C}^{1 \times NT},
\end{equation}
And similarly, concatenate the phase coefficients as:
\begin{equation}
\label{eq:e2}
\boldsymbol{v} = \begin{bmatrix} \boldsymbol{w}_1 \, \cdots, \ \boldsymbol{w}_T \end{bmatrix}^T \in \mathbb{C}^{NT \times 1}.
\end{equation}
Assuming the pilot is normalized and that the channel remains static over the T time slots, the total received power is:
$
P =
\sum_{t = 1}^{T} \left| \boldsymbol{\overline{h}(\theta)}^H \boldsymbol{w}_t \right|^2.
$. Thus, the optimization problem across multiple time slots can be rewritten as:
\begin{equation}
\label{eq:optmt}
\begin{aligned}
\max_{v_i \in \mathcal{S}, \forall i} \quad & \min_{\theta \in [\theta{\rm min},\theta{\rm max}]} \quad  \left| \boldsymbol{\overline{f}(\theta)}^H \boldsymbol{v} \right|^2.
\end{aligned}
\end{equation}

\section{Proposed Method}
\label{sec4}

In this section, we systematically addressed the RIS beam synthesis problem with discrete phase constraints. We began by reformulating the original optimization into an epigraph form to clarify the objective and constraints. Next, we relaxed the discrete phase constraint to its convex hull and introduced an $\ell^2$-norm penalty to encourage discrete solutions. To handle the resulting non-convexity, we employed the MM algorithm for efficient optimization. We also showed how this framework can be adapted to other practical constraints, such as constant modulus and per-element power constraints.

\subsection{$\Delta \theta$: Discretization of $\theta$}
In our approach to solving problem \eqref{eq:opt01}, a critical challenge is to handle the continuum of $\theta$ values in the ROI. To address this, we discretize the angle $\theta$ with a step size $\Delta\theta$. The discretization must be fine enough to capture the beam’s characteristics so that the power at all points within the region remains nearly the same, i.e., within a pre-specified bound. 

\begin{remark}
\label{prop:discretization}
For problem \eqref{eq:opt01}, the angle variable $\theta$ is continuous in ROI, hence a discretization with step size $\Delta\theta$ is required for numerical optimization. The role of $\Delta\theta$ is to make sure that the optimized wide beam remains flat when evaluated on the continuous ROI. Appendix \ref{appendix:discretization} provides a detailed analysis.
\end{remark}


 \vspace{-0.4cm}
\subsection{Handling Discrete Constraints via Convex Hull Relaxation}

In the previous step, we addressed the challenge of dealing with an infinite set of angle constraints by discretizing $\theta$ with an appropriate step size $\Delta\theta$. This allowed us to transform the original continuous optimization problem into a finite-dimensional form, making it computationally feasible.

However, another key challenge remains: the discrete phase constraints which make the problem NP-hard. In this step, we introduce an equivalent epigraph reformulation of the problem and employ a penalty-based convex hull relaxation. The goal is to transform the optimization problem into a convex-constraint problem while ensuring that the optimal solution remains within the discrete phase set.
To begin with, we rewrite the original problem (\ref{eq:opt01}) in its equivalent epigraph representation:
\begin{equation}
  \label{eq:opt3}
  \begin{aligned}
    \max_{w_i \in \mathcal{S}, \forall i} \quad & T^{\prime}\\
    \text{s.t.} \quad &  T^{\prime} \leq \min_{\theta \in [\theta_{min},\theta_{max}]} \quad  \left| \boldsymbol{\overline{h(\theta)}}^H \boldsymbol{w} \right|^2,
  \end{aligned}
\end{equation}
where $T^{\prime}$ is a dummy variable that represents the minimum power in the ROI. This formulation makes the constraint structure more explicit. 
Intuitively, the constraint in problem \eqref{eq:opt3} enforces that $T^{\prime}$ must be at most the smallest power achieved across the entire range of $\theta$. The objective is to maximize $T^{\prime}$, thereby ensuring the most uniform power distribution over the ROI.

\begin{figure}[t]
    \centering
    \subfloat[]{
        \centering
        \includegraphics[width=0.1\textwidth]{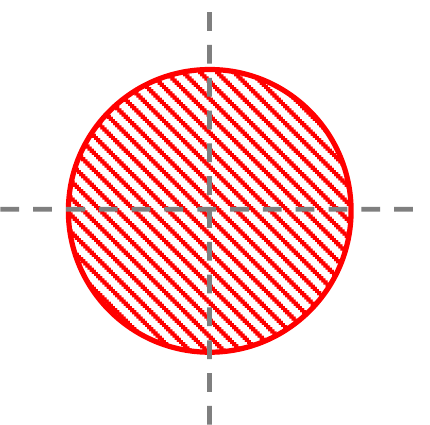}
        \label{fig:con1}}
        \hfill
    \subfloat[]{
        \centering
        \includegraphics[width=0.1\textwidth]{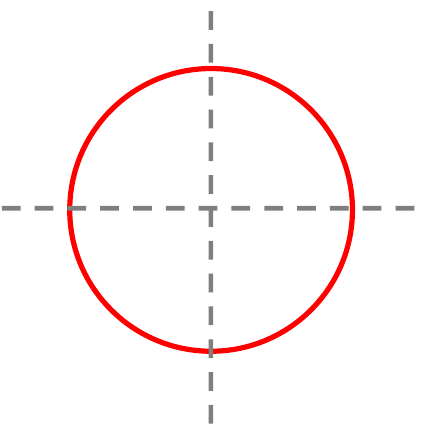}
        \label{fig:con2}}
        \hfill
   \subfloat[]{
        \centering
        \includegraphics[width=0.1\textwidth]{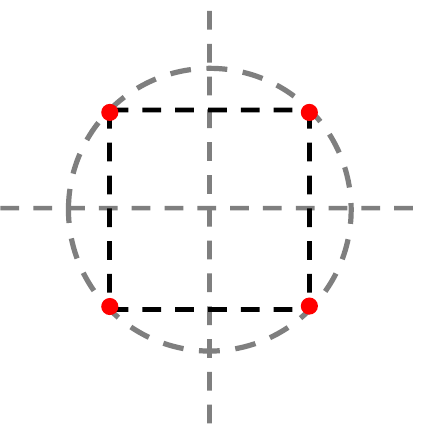}
        \label{fig:con3}}
        \hfill
    \subfloat[]{
        \centering
        \includegraphics[width=0.1\textwidth]{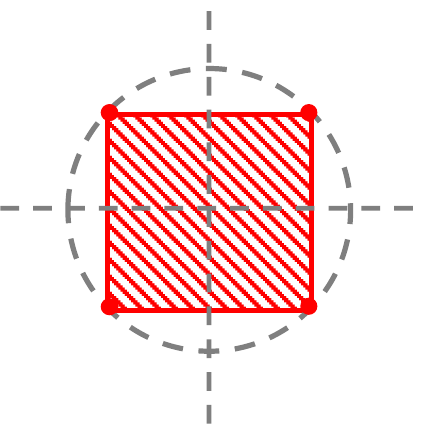}
        \label{fig:con4}}
    \caption{Illustration of constraints. (a) Per-element power constraint, (b) Constant modulus constraint (CMC), (c) Discrete phase constraint, (d) Convex hull of the discrete set.}
    \label{fig:convexhull}
        \vspace{-0.5cm}
  \end{figure}

Based on the epigraph representation, our next step is to relax the constraint from the discrete set $\mathcal{S}$ to its convex hull 
$
\mathcal{S}^{\prime} = \operatorname{conv}(\mathcal{S}).
$
This relaxation is illustrated in Fig.~\ref{fig:convexhull}, where:
Fig.~\ref{fig:con1} represents the per-element power constraint $ |w_i |^{2} \leq 1, \forall i \in N$, which allows possible absorption in RIS elements. Fig.~\ref{fig:con2} is the constant modulus constraint (CMC), as $ |w_i |^{2} = 1$, meaning that RIS elements can only adjust their phase. Figure \ref{fig:con3} is the discrete phase constraint $\mathcal{S}$ in Eq.~\eqref{eq:dpc}, which restricts the phase of each element to a finite set of values. Figure \ref{fig:con4} shows the smallest convex set  $\mathcal{S}^{\prime}$ that contains all the points of the discrete set.

To encourage solutions to the discrete set $\mathcal{S}$, we introduce an $\ell^2$-norm penalty term, controlled by a parameter $\lambda$, into the objective function.
With this modification, the optimization problem is reformulated as:
  \begin{equation}
    \label{eq:opt7}
    \begin{aligned}
      \max_{w_i \in \mathcal{S}^{\prime}, \forall i} \quad & T = T^{\prime} + \lambda \lVert \bm{w} \rVert _2 ^2 \\
      \text{s.t.} \quad & T \leq  f_\theta(\bm{w}) = \left| \boldsymbol{\overline{h(\theta)}}^H \boldsymbol{w} \right|^2 + \lambda \lVert \bm{w} \rVert_2^2, \\
       &  \forall \theta \in [\theta_{min},\theta_{max}] .
    \end{aligned}
  \end{equation}
Here, $T$ is bounded above by the function $ f_\theta(\bm{w})$, which includes both the signal power term and the penalty term.  

Given this setup, we now investigate whether the relaxed problem still yields a solution within the original discrete set $\mathcal{S}$. 
\textcolor{black}{To this end, we observe that the objective function $f_{\theta}(\bm{w}) = \big|\overline{h(\theta)}^H \bm{w}\big|^2$ is convex in $\bm{w}$ for each fixed $\theta$, and the added penalty term $\lambda\|\bm{w}\|_2^2$ preserves this convexity. Furthermore, all elements in the discrete phase set $\mathcal{S}$ share the same Euclidean norm. Hence, the conditions of the following proposition are satisfied in our setting:}
\begin{thm}
\label{pp1}
Let $\mathcal{S} = \{ s_1, \dots, s_L \} \subset \mathbb{C}$ be a discrete set of complex numbers, $|s_\ell| = \rho, \  \forall\ell$ and $\rho > 0$. Define the feasible set
\[
\mathcal{A}=\{\bm{x}\in\mathbb{C}^N:\; x_i\in\mathcal{S},\ i=1,\dots,N\},\] and let its convex hull be \[
\mathcal{A}'=\operatorname{conv}(\mathcal{A}).
\]
Let $f:\mathbb{C}^N\to\mathbb{R}$ be convex and $\lambda\ge0$. Then
\[
\argmax_{\bm{x}\in\mathcal{A}'}\ \big[f(\bm{x})+\lambda\|\bm{x}\|_2^2\big]
\;=\;
\argmax_{\bm{x}\in\mathcal{A}}\ \big[f(\bm{x})+\lambda\|\bm{x}\|_2^2\big].
\]
\end{thm}


The proof of Proposition \ref{pp1} can be found in Appendix \ref{app:p1}.

 \vspace{-0.4cm}
\subsection{Solving the Nonconvex Problem via the MM Algorithm}

Although the discrete constraint has been relaxed via the convex hull and an $\ell^2$-norm penalty has been introduced, the resulting problem in \eqref{eq:opt7} remains nonconvex due to the intrinsic structure of the objective function. 
\textcolor{black}{Specifically, in
$
f_\theta(\bm{w}) = \left| \overline{h(\theta)}^H \bm{w} \right|^2 + \lambda \|\bm{w}\|_2^2
$
is a convex quadratic function of $\bm{w}$. which makes maximizing it is generally a nonconvex task.} To efficiently address this nonconvexity, we employ the Minorization–Maximization (MM) algorithm \cite{sunMajorizationMinimizationAlgorithmsSignal2017}, which iteratively constructs and maximizes a sequence of tractable surrogate functions that lower-bound the original objective. The MM algorithm is implemented in two steps per iteration. 



 \begin{itemize}
  \item {Minorization Step:} A surrogate function $g_\theta(\bm{w}| \bm{w}^{(k)})$ is constructed at the current iterate $\bm{w}^{(k)}$ such that it lower bounds the original objective function $f_\theta(\bm{w})$. This surrogate is chosen to be easier to optimize.
  \item {Maximization Step:} The surrogate function is then maximized, yielding an updated solution $\bm{w}^{(k+1)}$.
  \end{itemize}

  

\textcolor{black}{For our particular case, recall that for any convex function $f(\bm{x})$, its first-order Taylor expansion at a reference point $\bar{\bm{x}}$ provides a global lower bound, i.e.,
\begin{equation}
    f(\bm{x}) \geq f(\bar{\bm{x}}) + \langle \nabla f(\bar{\bm{x}}), \bm{x} - \bar{\bm{x}} \rangle, \quad \forall \bm{x}.
\end{equation}}
In our setting, each term $f_{\theta}(\bm{w}) = \left| \overline{h(\theta)}^H \bm{w} \right|^2 + \lambda \|\bm{w}\|_2^2$ is convex and differentiable with respect to $\bm{w}$. Therefore, its first-order expansion around the current iterate $\bm{w}^-$ yields a global lower bound. Specifically, using $\|\bm{w}\|_2^2 \geq \|\bm{w}^-\|_2^2 + 2\langle \bm{w}^-, \bm{w} - \bm{w}^- \rangle$ and the same property for the quadratic term, we obtain the surrogate function:
\begin{equation}
\begin{aligned}
      \label{eq:lowbound}
    f_\theta(\bm{w}) & \geq   f_\theta(\bm{w}^-) + 2\langle \overline{\bm{h}}^H \bm{w}^-, \overline{\bm{h}}^H \bm{w} - \overline{\bm{h}}^H \bm{w}^- \rangle \\
      & \quad + 2 \lambda \langle  \bm{w}^-,  \bm{w} - \bm{w}^- \rangle\\
      & = g_\theta(\bm{w}| \bm{w}^-), \quad \forall \theta \in [\theta_{\min},\theta_{\max}],
\end{aligned}
\end{equation}
which holds globally for all $\bm{w}$ and $\theta$.

With this surrogate, the MM algorithm updates the solution at each iteration by solving the following convex problem:
\begin{equation}
  \label{eq:opt8}
  \begin{aligned}
  \max_{ w_i \in \mathcal{S}^{\prime}, \forall i} \quad & T \\
  \text{s.t.} \quad & T \leq g_\theta(\bm{w}| \bm{w}^{(k)}), \quad \forall \theta \in [\theta_{\min},\theta_{\max}].
  \end{aligned}
  \end{equation}
$T$ is an auxiliary variable representing the lower bound on the achieved objective over the entire angle range. The solution $\bm{w}^{(k+1)}$ obtained from \eqref{eq:opt8} is then used for the next MM step.

Given the iterative nature of this approach, it is essential to establish the convergence of the algorithm. This brings us to the following proposition:
\begin{thm}
\label{pp2}
The proposed algorithm is guaranteed to converge to a stationary point of the MM algorithm.
\end{thm}
The proof of Proposition \ref{pp2} is shown in Appendix \ref{app2}.
    

  \textcolor{black}{Once the MM algorithm converges, the resulting solution $\bm{w}^{(k+1)}$ generally lies on the boundary of the convex hull $\mathcal{S}'$. Although Proposition~\ref{pp1} guarantees that the global optimum of the relaxed problem lies in the original discrete set $\mathcal{S}$, the MM algorithm is only guaranteed to converge to a stationary point. As a result, the iterate $\bm{w}^{(k+1)}$ may not exactly coincide with an element of $\mathcal{S}$. To ensure adherence to the original discrete phase constraint, we perform an element-wise projection to $\mathcal{S}$:}
\begin{equation}
    \mathcal{P}(w_i) = \arg\min_{s \in \mathcal{S}} \left| \angle (w_i) - \angle (s) \right|,
\end{equation}
where $ \angle (w_i) = \frac{w_i}{|w_i|}$ returns the phase angle of $ w_i $. 
The operator $\mathcal{P}$ rounds the phase of each element in $ \bm{w} $ to the nearest phase in the discrete set $ \mathcal{S} $.
The overall procedure is summarized in Algorithm~\ref{alg:MM2}.

\begin{algorithm}[htb]
  \caption{MM method for Problem (\ref{eq:opt7}) via a penalty method}
  \label{alg:MM2}
  \begin{algorithmic}[1]
    \STATE Initialize $\boldsymbol{w}^{(0)}$ randomly, 
    \REPEAT
    \STATE Find the surrogate function $g_\theta(\bm{w}| \bm{w}^{(k)})$ as in Eq.~(\ref{eq:lowbound}).
    \STATE Solve the convex problem:
    \begin{equation}
      \label{eq:opt11}
      \begin{aligned}
        \max_{\bm{w} \in \mathcal{S}^{\prime}} \quad &  T \\
    \text{s.t.} \quad &  T \leq g_\theta(\bm{w}| \bm{w}^{(k)}), \forall \theta \in [\theta_{min},\theta_{max}].
      \end{aligned}
    \end{equation}
    \STATE Let $\boldsymbol{w}^{(k+1)}$ be the solution obtained in Step 4.
    \UNTIL{ $T(\bm{w}^{(k)})$ - $T(\bm{w}^{(k+1)})< \epsilon $.} 
    \STATE Output $\bm{w}^* = \mathcal{P}(\bm{w}^{(k+1)})$.
  \end{algorithmic}
\end{algorithm}
\vspace{-3\lineskip}

\begin{figure*}[t]
  \centering
  
  \begin{subfigure}[b]{0.32\textwidth}
      \centering
      \includegraphics[width=\textwidth]{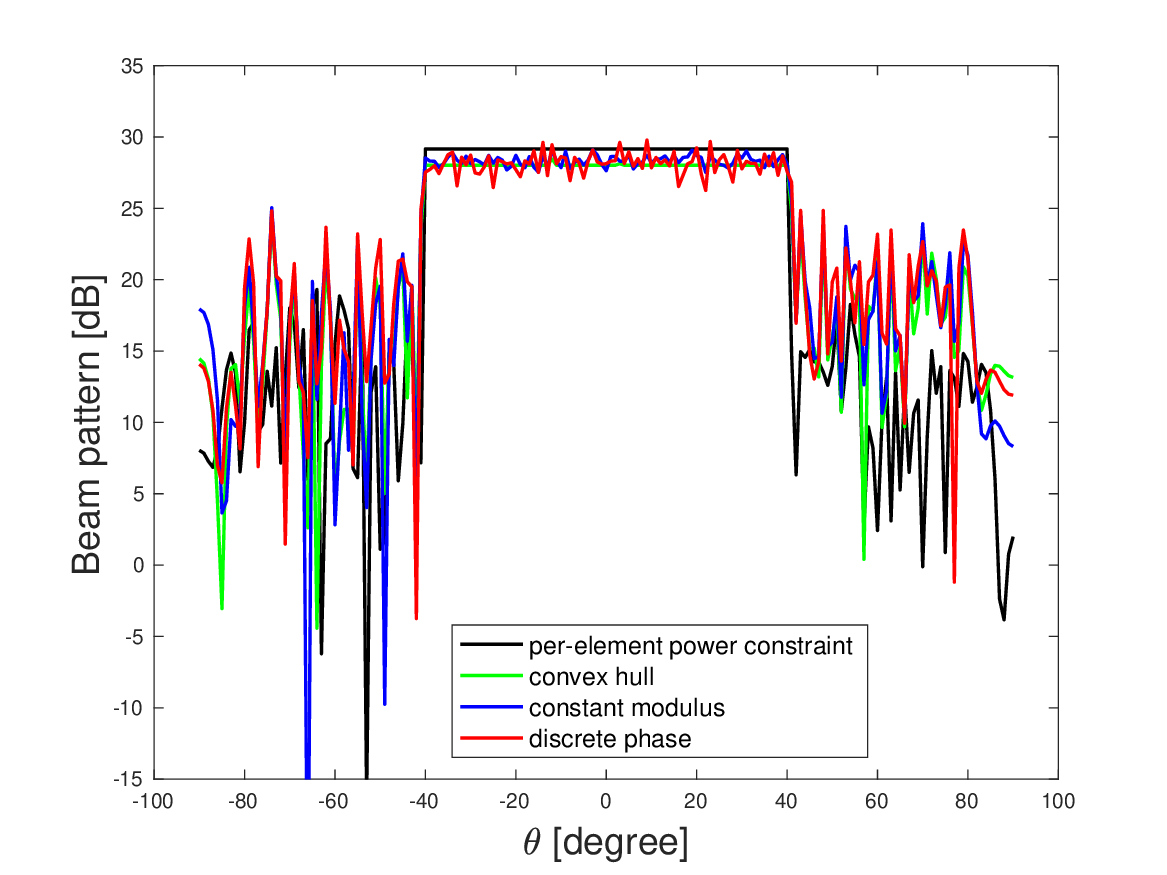}
      \caption{Interest region: $[-40^\circ, 40^\circ]$}
      \label{fig:tar2}
  \end{subfigure}
  \begin{subfigure}[b]{0.32\textwidth}
      \centering
      \includegraphics[width=\textwidth]{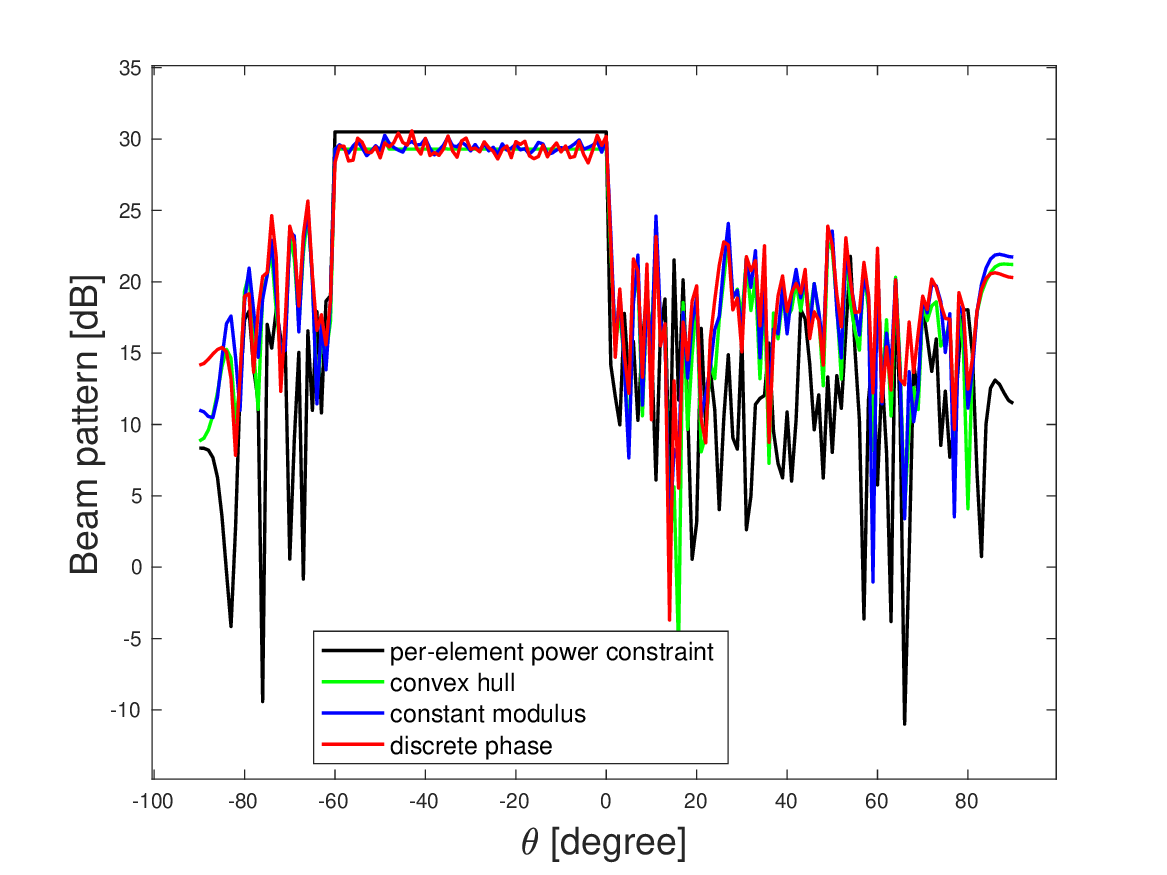}
      \caption{Interest region: $[-60^\circ, 0^\circ]$}
      \label{fig:tar3}
  \end{subfigure}  
  \begin{subfigure}[b]{0.32\textwidth}
      \centering
      \includegraphics[width=\textwidth]{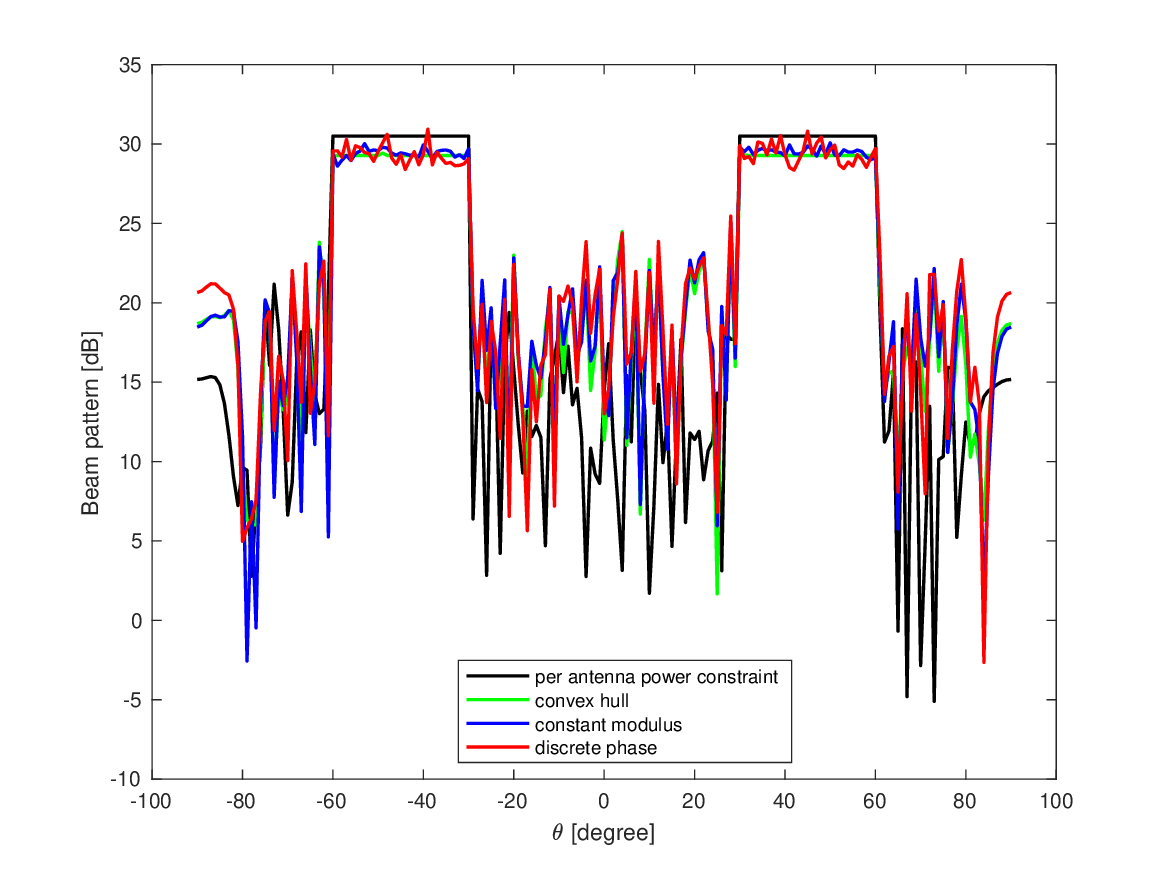}
      \caption{Interest region: $[-60^\circ, -30^\circ] \cup [30^\circ, 60^\circ]$}
      \label{fig:2beam}
  \end{subfigure}
  \caption{Beam pattern for the target at different locations. Parameters: $N =256$, $L = 4$.}
  \label{fig:tartar}
  \vspace{-0.6cm}
\end{figure*}

 \vspace{-0.4cm}
\subsection{Adaptation to different constraints}
\subsubsection{CMC}  Our algorithm can also be applied to scenarios with constant modulus constraints. In such cases, our algorithm reduces to the classical penalty method \cite{bertsekas2003convex}. The projection operator $\mathcal{P}_{\text{CMC}}$ is utilized to retain only the phase of each $w_i$, ensuring that the modulus remains constant, like Fig.~\ref{fig:con3}.To enforce this constraint after the MM iterations, we apply a projection operator for the CMC, which is:
\begin{equation}
    \mathcal{P}_{\text{CMC}}(w_i) = \angle (w_i).
\end{equation}

\subsubsection{Per-element power constraint}
In some scenarios, the RIS weights may only be required to satisfy a per-element power constraint, namely $|w_i|^2 \leq 1$, the problem is:
\begin{equation}
  \label{eq:opt6}
  \begin{aligned}
    \max_{} \quad & T \\
    \text{s.t.} \quad &  |w_i|^2 \leq 1, \quad i=1,\ldots,N\\
    & T \leq f_\theta(\bm{w}), \quad \forall \theta \in [\theta_{min},\theta_{max}] .
  \end{aligned}
\end{equation}
 \vspace{-0.4cm}

\section{Beam pattern simulation results}
\label{secbmsim}

Building on the formulations and algorithmic developments presented above, we now illustrate the effectiveness of the proposed beam synthesis method through beam pattern analysis.

We consider an RIS with different numbers of elements, and the distance between adjacent elements is $0.5\lambda$. Assume that the BS is located at $\phi = 0^\circ$. The convergence threshold is set at $\epsilon = 10^{-4}$.
\subsubsection{ULA simulation results}
First,  we verify the results of our algorithm by comparing them with the results of different constraints. Figure \ref{fig:tartar} plots the results for different target locations. We can see that the minimum power in the ROI is high enough to distinguish it from the side region. Furthermore, each subfigure presents 4 different constraints, namely, per-element power constraint, convex hull, constant modulus, and discrete phase ($L =4$), which is shown in Fig.~\ref{fig:convexhull}. It can be observed that, with per-element power constraints, the beam pattern is the flattest, since it has the least stringent constraints. In the other three cases, the beam pattern fluctuates more as the constraints become stricter. The most pronounced fluctuations are observed in the case of the discrete phase with a quantization level of $L = 4$. However, it remains comparable to that under the per-element power constraint, demonstrating that our method can generate a wide beam, both for continuous and discrete phase coefficients.

\begin{figure*}[t]
  \centering
  \begin{subfigure}[b]{0.32\textwidth}
      \centering
      \includegraphics[width=\textwidth]{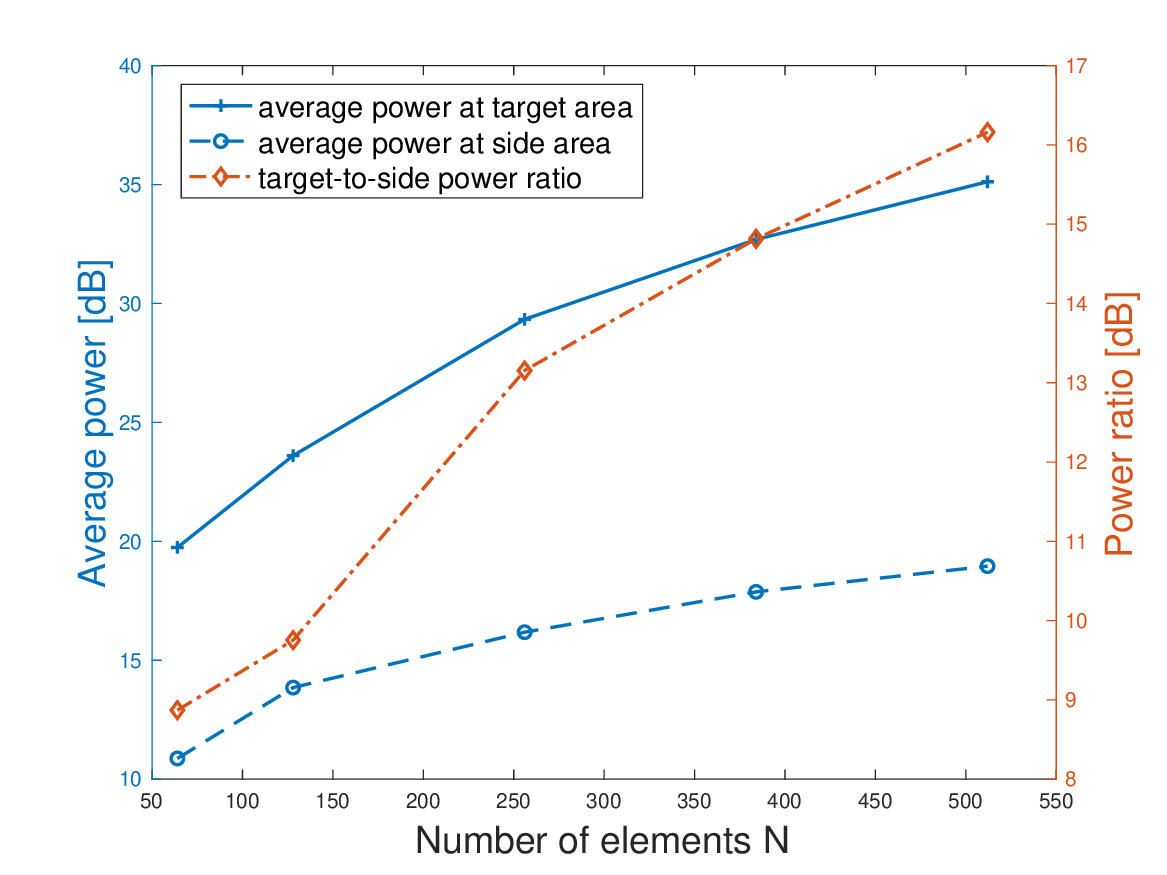}
      \caption{Average power at target region and side region for different $N $, $L = 4$.}
      \label{fig:N}
  \end{subfigure}
  \begin{subfigure}[b]{0.32\textwidth}
      \centering
      \includegraphics[width=\textwidth]{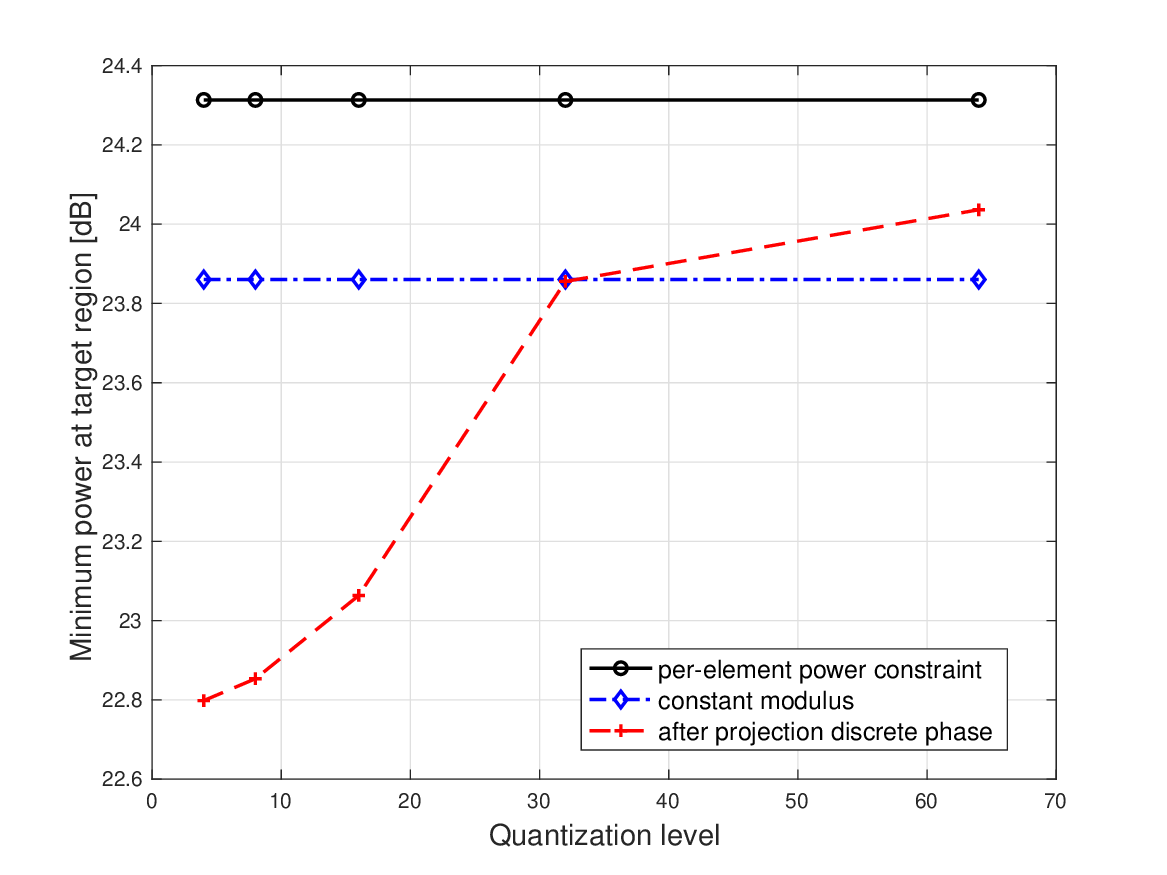}
      \caption{Minimum power with different quantization level $L$, $N = 128$.}
      \label{fig:l}
  \end{subfigure}  
  \begin{subfigure}[b]{0.32\textwidth}
      \centering
      \includegraphics[width=\textwidth]{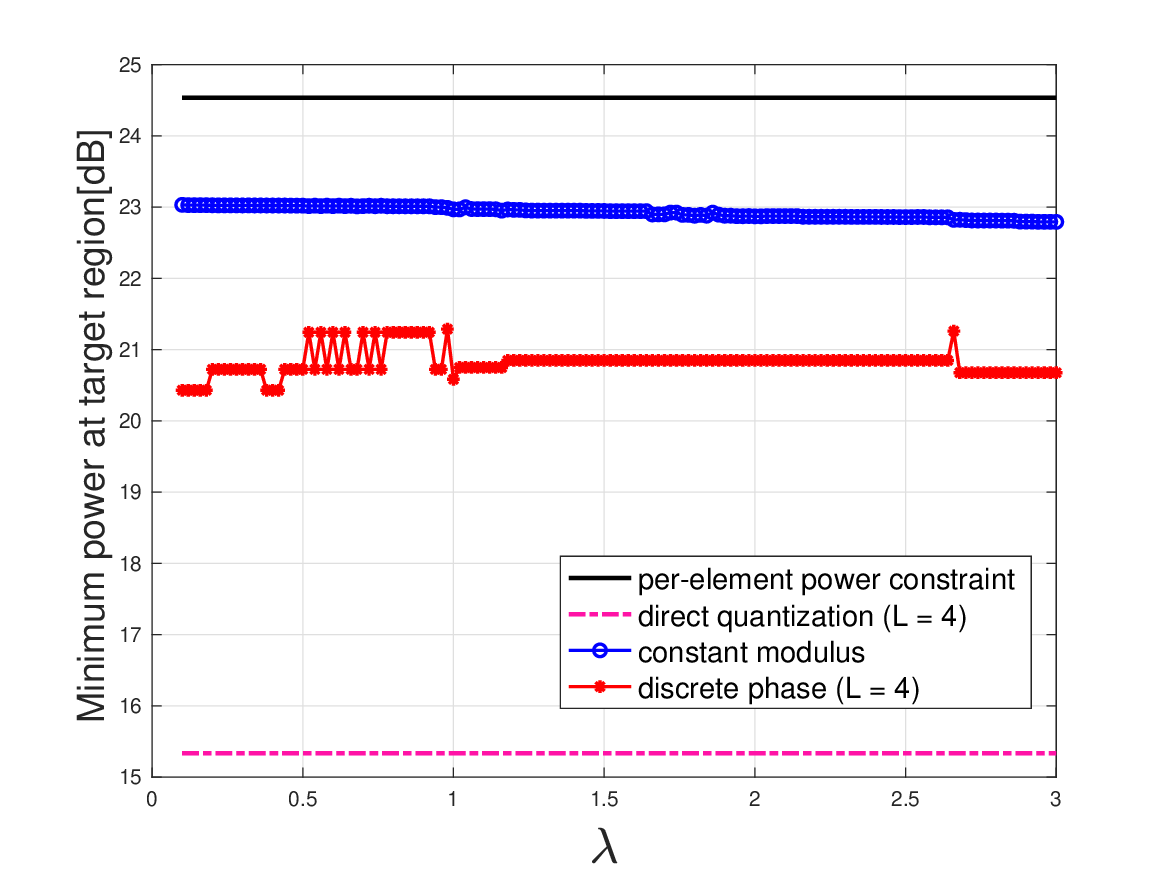}
 \caption{The effect of penalty term $\lambda$. Parameters: $N = 128$, $L = 4$.} 
 \label{fig:lambda}
  \end{subfigure}
  \caption{The effect of different parameters. Target region at $[-30^\circ, 30^\circ]$.}
  \label{fig:ulaall}
    \vspace{-0.6cm}
\end{figure*}

\textcolor{black}{Figure \ref{fig:N} shows the effect of the number of elements $N$. We can see that the minimum power at both the target region and the side region increases as $N$ increases. For a fully coherent array configuration where all elements are aligned toward the same direction, the received power scales proportionally to beamforming gain $N^2$. 
Although exact phase alignment is not achievable in wide-beam synthesis, the dominant trend still holds: increasing $N$ enhances the overall array gain, which translates into higher minimum power across the angular ROI.}


\begin{figure}[htb]
\captionsetup[subfigure]{skip=-10pt}   
  \centering
   \begin{subfigure}[b]{0.4\textwidth}
      \centering
      \includegraphics[width=\textwidth, trim={0 5bp 0 0},clip]{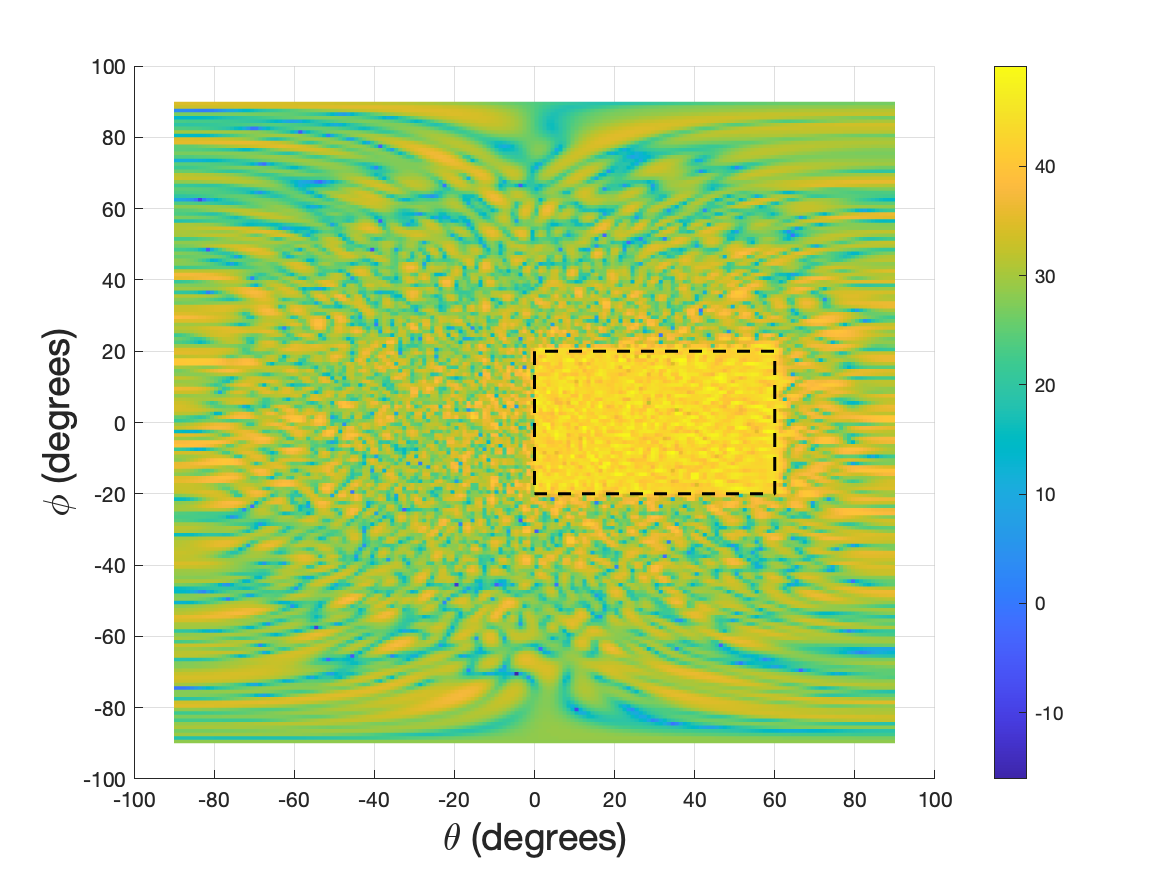}
      \label{fig:upa1}
      \caption{Target region: $\theta \in [0^\circ, 60^\circ]$, $\phi \in [-20^\circ, 20^\circ]$}
  \end{subfigure}
  \begin{subfigure}[b]{0.4\textwidth}
      \centering
      \includegraphics[width=\textwidth, trim={0 5bp 0 0},clip]{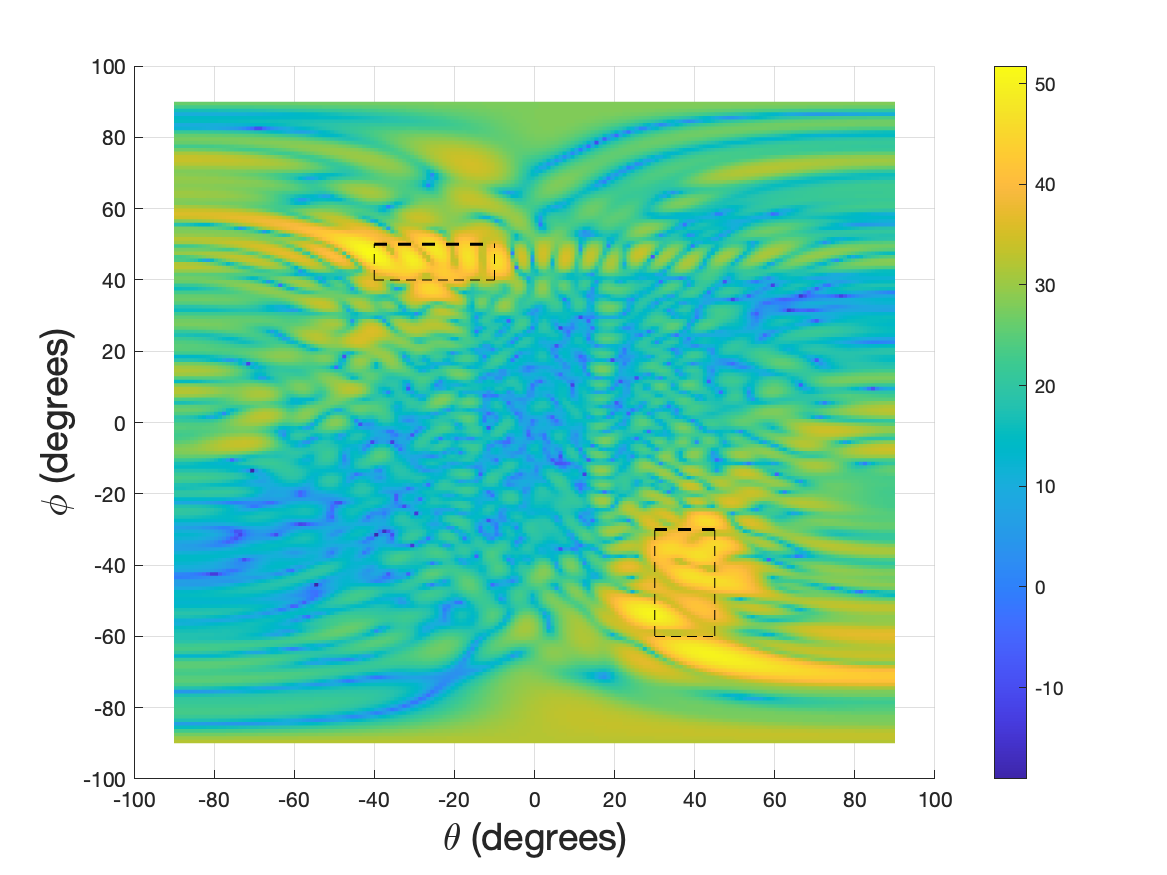}
      \label{fig:upa2}
      \caption{Target region: $\theta \in [-40^\circ, -10^\circ]$, $\phi \in [40^\circ, 50^\circ]$ and $\theta \in [30^\circ, 45^\circ]$, $\phi \in [-30^\circ, -60^\circ]$}
  \end{subfigure}  
  \caption{Beam pattern for a UPA with $N = 64\times64$, $L =4$.}
  \label{fig:upa}
\vspace{-0.5cm}
\end{figure}

Figure \ref{fig:l} shows the effect of different quantization levels $L$. We observe that the minimum power at the target region increases as the quantization level increases. This is because the discrete phase set becomes larger as the quantization level increases. However, the performance gain is not significant as the quantization level increases further. This is because the beam pattern is already achievable enough when $L = 4$. Another observation is that when $L = 64$, the minimum power in the target region in the discrete phase case exceeds that of the constant modulus. This is because neither of them is a globally optimal solution, and the constant-modulus solution is not necessarily better.

Figure \ref{fig:lambda} shows the effect of the penalty term $\lambda$. We compare the minimum power at the target region for different constraints and different values of $\lambda$. The best case is the per-element power constraint, which aligns with the beampattern shown in Fig.~\ref{fig:tartar}. The worst case, which is obtained through direct quantization of the per-element power constraint solution, exhibits around a 10dB degradation compared to the per-element power constraint. The CMC and the discrete phase solution of our method achieve better results than the direct quantization. Our solution is around 5dB better than the direct quantization. Also, when the penalty term $\lambda$ is large, the performance tends to stabilize, verifying Proposition \ref{pp1}.

\subsubsection{UPA simulation result}
In Fig.~\ref{fig:upa}, we show the performance of our method applied to a UPA with $N = 64\times64$ elements. $\theta$ and $\phi$ are the azimuth and elevation angles, respectively. The distance between adjacent elements is still $0.5\lambda$. The beam pattern matches the target region which indicates that our algorithm can be applied to a UPA as well.

  

\section{APPLICATION in AOA estimation}
\label{sec5}

\begin{figure}[tb]
    \centering
    \begin{subfigure}[b]{0.225\textwidth}
        \centering
        \includegraphics[width=\textwidth]{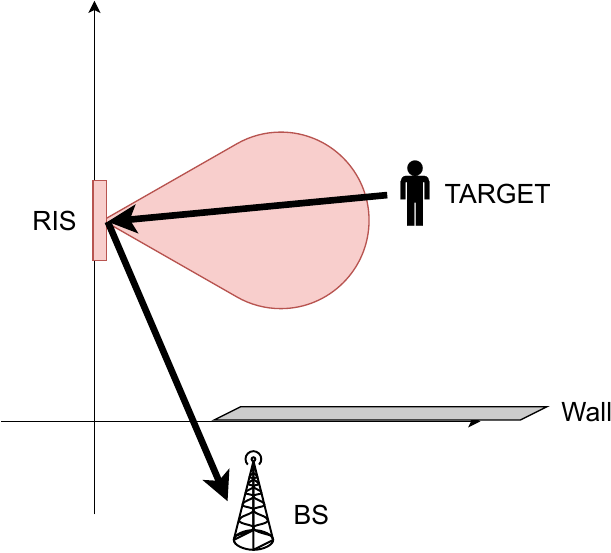}
        \caption{Beam sweeping method}
        \label{fig:sweep}
    \end{subfigure}
    \hfill
    \begin{subfigure}[b]{0.225\textwidth}
        \centering
        \includegraphics[width=\textwidth]{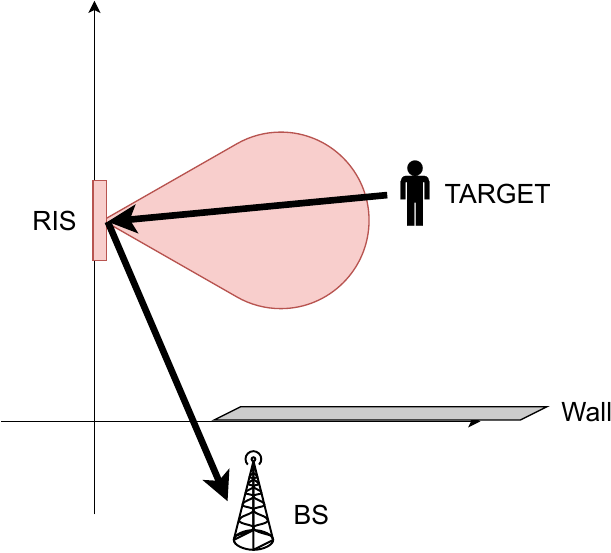}
        \caption{Wide beam method}
        \label{fig:wide}
    \end{subfigure}
    \caption{Comparison between the beam sweeping method and the wide beam method.}
    \label{fig:comparison}
  \end{figure}

In this section, we explore the application of our proposed beam synthesis method in angle-of-arrival (AOA) estimation. The traditional approach is to use a \textit{beam sweeping} method, \textcolor{black}{in which the base station sequentially transmits a series of narrow beams, each pointed in a different direction.} The received signals are analyzed to determine the target's direction. Although effective, beam sweeping requires a large number of pilots to cover a wide angular region, especially when the area of interest is broad. In contrast, our proposed wide beam method directly synthesizes a single beam that uniformly covers the entire region, reducing the number of required pilots. Figure~\ref{fig:comparison} highlights the difference between these two approaches: the beam sweeping method generates a series of narrow beams to cover the target area, our method employs a single wide beam to cover the same area in one step.

In the following, we first detail the temporal-domain MUSIC algorithm for AOA determination, which serves as the basis of our estimation framework. Building on this foundation, we present numerical results that demonstrate the performance of our wide beam method in AOA estimation under various simulation settings.

\subsection{Temporal-domain MUSIC algorithm}
The Multiple Signal Classification (MUSIC) algorithm \cite{stoicaMUSICMaximumLikelihood1989} is a classic spectral estimation method used to determine the angle of arrival (AOA) of multiple signals impinging on an array of sensors. This technique is based on the decomposition of the signal space into orthogonal subspaces corresponding to the signal and noise components.
Assume an array of $M$ antennas receiving signals from $K$ far-field sources. Denote $\theta_k$ as the AOA from the source $k$. The received signal vector $ \bm{x}_t $ at any discrete time $ t $ can be expressed as:
$
    \bm{y}_t = \sum_{k=1}^{K} \bm{a}(\theta_k) s_{k,t}  + n_t = \bm{A}(\bm{\theta}) \bm{s}_t + \bm{n}_t.
$
    The covariance matrix $ \bm{R} $ of the received signal is the average the outer product of $ \bm{y}_t $ over $ T $ samples:
    $
    \bm{R} = \frac{1}{T} \sum_{t=1}^T \bm{y}_t \bm{y}_t^H.
    $

   Perform the eigendecomposition of the covariance matrix $ \bm{R} $:
    $
    \bm{R} = \bm{E}_s \bm{\Lambda}_s \bm{E}_s^H+ \bm{E}_n \bm{\Lambda}_n \bm{E}_n^H,
   $
 and \( \bm{E}_s \) and \( \bm{E}_n \) are the matrices containing the eigenvectors corresponding to the signal and noise subspaces, respectively, and \( \bm{\Lambda}_s \), \( \bm{\Lambda}_n \) are diagonal matrices of their associated eigenvalues. Then, we can apply the traditional MUSIC algorithm by computing the MUSIC spectrum: \begin{equation}
         P_{\text{MUSIC}}(\theta) = \frac{\bm{A}(\theta)^H\bm{A}(\theta)}{\bm{A}(\theta)^H \bm{E}_n \bm{E}_n^H \bm{A}(\theta)}
    \end{equation}   and the AOA estimates are identified by locating the $K$ peaks in the MUSIC spectrum \( P_{\text{MUSIC}}(\theta) \).

However, if we investigate our channel model from Eq.~\eqref{eq:sysmodel} in its steering vectors, we see that
$ \bm{a}(\phi)^H \text{diag}(\bm{w}) \bm{a}(\theta) \in \mathbb{C}^{1\times1}$, and it does not provide independent observations in the spatial domain. Inspired by \cite{liuliangmusic2023},
we can create several independent observations in the time domain by changing the reflective pattern of RIS. To generate independent observations in the time domain, we vary the RIS phase configuration over $T$ time slots. We follow the same setting as in our previous formulation (see Eq.~\eqref{eq:sysmodeltotal}). In the time-domain approach, by varying the RIS phase configuration over $T$ time slots, the received signal can be expressed as:
$
    y_{t} = \boldsymbol{\overline{h}}^H \boldsymbol{w}_{t} s + n_{t}, \ t=1,\ldots,T,
$
And by stacking these observations, we have
\begin{equation}
    \bm{\bar{y}} = \boldsymbol{\overline{h}}^H \boldsymbol{W} s + \bm{\bar{n}} \in \mathbb{C}^{1\times T},
\end{equation}
where $\boldsymbol{W}$ is defined as 
$\boldsymbol{W} = \left[ \boldsymbol{w}_{1}, \boldsymbol{w}_{2},\cdots, \boldsymbol{w}_{T}  \right] \in \mathbb{C}^{N\times T}$ and the additive noise is $\bm{\bar{n}} = [n_1, n_2, \cdots, n_T]$.

\textcolor{black}{To obtain sufficient snapshots for subspace-based estimation techniques such as MUSIC, the same set of RIS configurations $\boldsymbol{W}$ is repeated over $Q$ independent blocks.} In each block $q \in \{1,\ldots,Q\}$, a noisy copy of $\bm{\bar{y}}$ is collected, resulting in $Q$ independent observations $\{ \bm{\bar{y}}^{(q)} \}$. The sample covariance matrix is then estimated as
\begin{equation}
\bm S = \frac{1}{Q} \sum_{q=1}^{Q} \left( \bm{\bar{y}}^{(q)} \right)^H  \bm{\bar{y}}^{(q)} \in \mathbb{C}^{T \times T}.
\end{equation}

\textcolor{black}{The number of blocks \(Q\) controls the statistical accuracy of the covariance estimate. In general, a larger \(Q\) yields a more accurate estimate. However, this improvement comes at the cost of increased pilot overhead. 
In this work, we follow the setup in \cite{liuliangmusic2023}, which shows that \(Q = 4\) already achieves reliable estimation performance in moderate-SNR scenarios.}


Then, define the eigenvalue decomposition (EVD) of $ \bm S $ as
$ \bm S = \bm U \bm \Lambda \bm U^H $, where $ \bm \Lambda = \text{diag}(\lambda_1, \ldots, \lambda_L) $ whose diagonal elements are the eigenvalues of $ \bm S $, and $ U = [u_1, \ldots, u_T] $
consists of the corresponding eigenvectors. Since we only have one object, define
$ \bar{U} = [u_{2}, \ldots, u_T] \in \mathbb{C}^{T \times (T-1)} $. Therefore, the spectrum
used in the MUSIC algorithm is defined as
\begin{equation}
P(\theta) = \frac{\bm{\bar{H}}(\theta)^H \bm{\bar{H}}(\theta)}{\bm{\bar{H}}(\theta)^H \bar{U} \bar{U}^H \bm{\bar{H}}(\theta)}, \quad \forall \theta,
\end{equation}
where $\bm{\bar{H}}(\theta) = \boldsymbol{\overline{F}(\theta)}^H \boldsymbol{W}$.
Thus, we can perform a one-dimensional search to find the AOA. 


\subsection{Numerical results}
In this subsection, we aim to simulate the performance of our wide beam method in AOA estimation. 
The simulation settings are: the BS has a signal antenna, and the number of antennas at the RIS is 64,128.
The ROI is $ \theta \in [ -30^\circ, 30^\circ] $. 
The time sample $T$ is set to 7. The number of blocks used is set to $Q =4$. 
Within each block, the RIS phase configuration is chosen differently from the results of our proposed wide beam method provided by a different $\lambda$; then, the same configuration is repeated for the other block.

In the beam sweeping method, the RIS phase configuration is chosen in a sweeping manner, using the method in \cite{ren2022linear} to generate a beam pattern corresponding to 7 possible angles in the ROI. 
We set the object uniformly distributed in the ROI and run 5000 Monte Carlo simulations to generate random objects within $[-30^{\circ},30^{\circ}]$.
\textcolor{black}{This means the angular spacing between the sweeping beams is 10 degrees.}
The music algorithm searches the entire ROI with an accuracy of $10^{-4}$.

\begin{figure}[tb]
    \centering
    \includegraphics[width=0.47\textwidth]{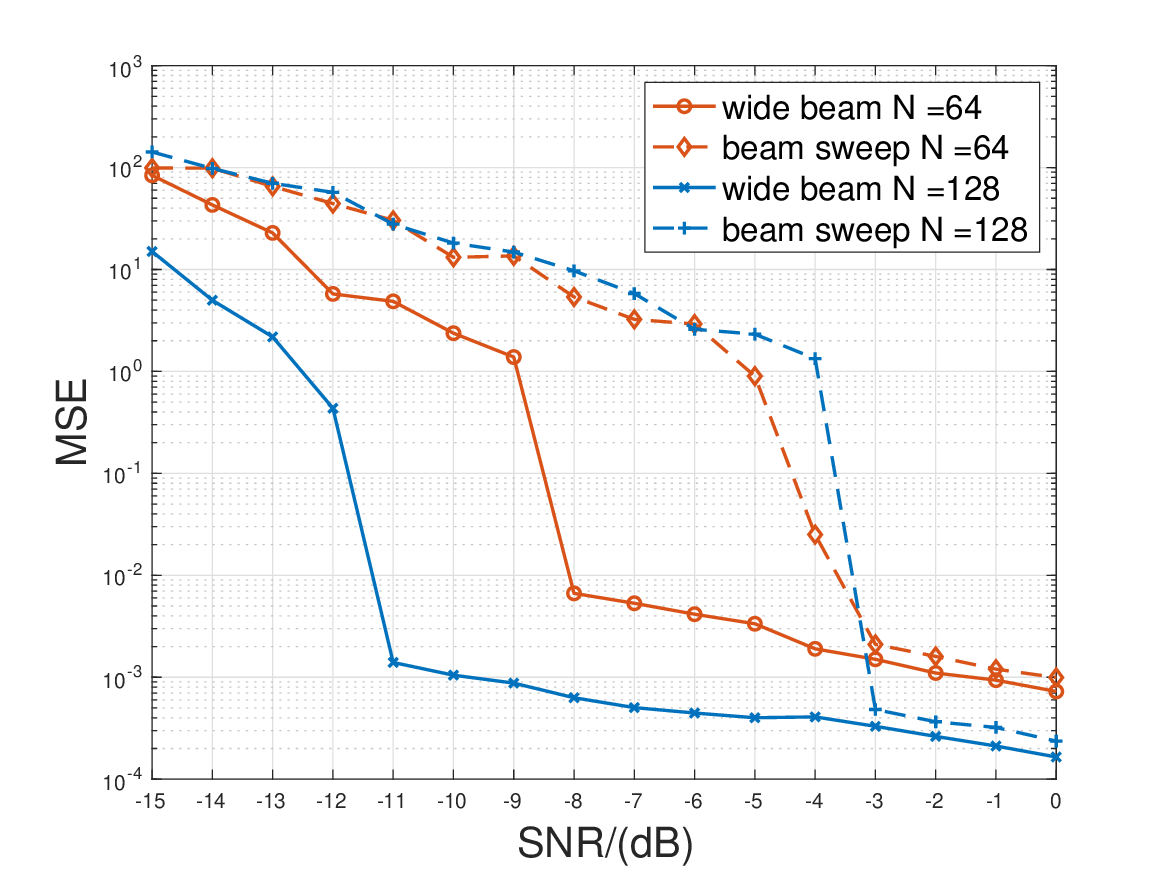}
    \caption{MSE (degree$^2$) of AOA estimation results of the beam sweeping method and the proposed method.}
    \label{fig:mse}
  \end{figure}

\textcolor{black}{
The MSE performance of the two AOA estimation methods is shown in Fig.~\ref{fig:mse}. Across the entire SNR range, the proposed wide beam method consistently outperforms beam sweeping for both RIS sizes. To reach an MSE of \(10^{-2}\), the wide beam approach requires around 5 dB and 8 dB less transmit power than beam sweeping for \(N=64\) and \(N=128\), respectively. At high SNR, the performance gap narrows since the signal subspace becomes dominant, allowing MUSIC to accurately resolve the target angle even with suboptimal beam patterns.}

\textcolor{black}{
A key observation from Fig.~\ref{fig:mse} is that beam sweeping does \emph{not} improve as the RIS size increases from \(N=64\) to \(N=128\); rather, its performance degrades at low SNR. This occurs because, for larger arrays, the main lobes become narrower, and many target angles fall outside the beam coverage, resulting in severe power loss. In contrast, the proposed wide beam ensures uniform power across the entire region and remains robust as \(N\) increases. This confirmsthat the proposed method not only achieves higher estimation accuracy but also scales more effectively with RIS size than beam sweeping.
}
 


\section{Application in Target Detection}

\subsection{GLRT formulation}
\textcolor{black}{
Target detection with unknown parameters leads to a joint detection and estimation problem, where the classic likelihood ratio test (LRT) is inapplicable. The generalized likelihood ratio test (GLRT) overcomes this by replacing unknown parameters with their estimates. Although this introduces estimation error, the GLRT is asymptotically equivalent to the optimal LRT as the number of observations grows \cite{DetectionEstimationModulation}. Here, we adopt the GLRT framework due to its balance between performance and computational efficiency.}

In our setting, the received BS during $T$ time slots is:
\begin{align}
  \mathcal{H}_0: &\quad y_t = n_t, \quad t = 1,\ldots,T, \\
  \mathcal{H}_1: & \quad y_t = \boldsymbol{H}_t(\theta) s + n_t, \quad t = 1,\ldots,T,
\end{align}
where $\mathcal{H}_0$ and $\mathcal{H}_1$ denote the target-absent and target-present hypotheses, respectively; $s$ is the known pilot, $\boldsymbol{H}_t(\theta)$ is the RIS-assisted channel parametrized by the unknown angle $\theta$, and $n_t$ is additive complex Gaussian noise with variance $\sigma^2$ (normalized to 1). The key challenge is that under $\mathcal{H}_1$, the parameter $\theta$ is unknown and must be estimated from the data.

When unknown parameters are present under the alternative hypothesis, the classical likelihood ratio test (LRT) is not directly applicable since the exact likelihood $\mathcal{L}(y|\mathcal{H}_1)$ is unavailable. In such cases, the generalized likelihood ratio test (GLRT) is widely adopted \cite{DetectionEstimationModulation}. The GLRT replaces unknown parameters with their maximum likelihood (ML) estimates and compares:
\begin{equation}
  \Lambda(\bm{y}) = \frac{\max_{\theta}~\mathcal{L}(\bm{y}~|~\mathcal{H}_1,\theta)}{\mathcal{L}(\bm{y}~|~\mathcal{H}_0)} 
\quad \mathop{\gtrless}_{\mathcal{H}_0}^{\mathcal{H}_1} \quad \eta,
\end{equation}
where $\bm{y} = \left[ y_{1} \cdots y_T \right]^T $ is the concatenated received signal within $T$ time samples. $\eta$ represents the decision threshold.



Because the noise samples are i.i.d. circularly symmetric complex Gaussian with normalized variance, the probability density function (PDF) of the observation vector $\bm{y}$ is
\begin{equation}
    f(\bm{y}) = \frac{1}{\pi^{T} \det(\bm{R}) } \exp{\left(-(\bm{y}- \bm{\mu})^H \bm{R}^{-1}(\bm{y}-\bm{\mu}) \right)},
\end{equation}
where $\bm{R}$ denotes the covariance matrix, and $\bm{\mu}$ is the mean vector under the corresponding hypothesis.

Under $\mathcal{H}_0$, the mean is zero. Under $\mathcal{H}_1$, the signal enters as a deterministic mean shift. In both cases, since the channel and signal are deterministic and the noise is white, the covariance matrix reduces to $\bm{R} = \sigma^2 \bm{I}_T = \bm{I}_T$. Substituting this into the expression above yields
\begin{equation}
\begin{aligned}
f(\bm{y} \mid \mathcal{H}_0) &= \frac{1}{\pi^{T}} \exp{\left( - \lVert \bm{y} \rVert^2 \right)},\\
f(\bm{y} \mid \mathcal{H}_1) &= \frac{1}{\pi^{T}} \exp{\left( - \lVert \bm{y} - \bm{\mu}(\theta) \rVert^2 \right)},
\end{aligned}
\end{equation}
where $\bm{\mu}(\theta) = \left[ \boldsymbol{H}_1(\theta)s, \cdots, \boldsymbol{H}_T(\theta)s \right]^T$ is the mean under $\mathcal{H}_1$. Substituting these PDFs into the GLRT expression yields a test statistic that depends on the unknown parameter $\theta$.

\textcolor{black}{Ideally, the GLRT requires the maximum likelihood estimate (MLE) of the unknown angle $\theta$ under $\mathcal{H}_1$. However, solving for the MLE is computationally intensive when the parameter space is continuous. To reduce complexity, we replace the MLE with the MUSIC estimator introduced in the previous section. It has been proved in \cite{stoicaMUSICMaximumLikelihood1989} that the MUSIC estimator is a large sample realization of the MLE and  it is asymptotically equivalent to the MLE at high SNR. This makes the detector both practical and well aligned with our RIS-based wide-beam sensing framework.}

Denote the MUSIC estimate as $\hat{\theta}$,
we formulate the GLRT as:
\begin{equation}
    \Lambda(\bm{y} | \hat{\theta} ) = \frac{e^{-\lVert \bm{y}-\bm{\hat{\mu}}\lVert ^2 }}{e^{-\lVert \bm{y}\lVert ^2}}.
\end{equation}
Take the logarithm on both sides, we can get:
\begin{equation}
    \begin{aligned}
        \log \Lambda(\bm{y}) &= -\lVert \bm{y}-\bm{\hat{\mu}}\lVert ^2 + \lVert \bm{y}\lVert ^2 \\
        & = -\lVert \bm{y}\lVert ^2 + 2 \Re \{ \bm{y}^H \bm{\hat{\mu}} \} - \lVert \bm{\hat{\mu}}\lVert ^2 + \lVert \bm{y}\lVert ^2 \\
        & =  2 \Re \{ \bm{y}^H \bm{\hat{\mu}} \} - \lVert \bm{\hat{\mu}}\lVert ^2 \mathop{\gtrless}_{\mathcal{H}_0}^{\mathcal{H}_1} \log \eta,
    \end{aligned}
\end{equation}
which corresponds to:
\begin{equation}
   \Lambda^{\prime} = \Re \{ \bm{y}^H \bm{\hat{\mu}} \}  \mathop{\gtrless}_{\mathcal{H}_0}^{\mathcal{H}_1} \frac{\log \eta +\lVert \bm{\hat{\mu}}\lVert ^2}{2} = \eta^{\prime}.
\end{equation}

The performance of the detector can be evaluated by the probability of false alarm and the probability of detection. The probability of a false alarm is given by:
\begin{equation}
    P_{\text{FA}} = \mathbb{P} \left( \Lambda^{\prime} > \eta^{\prime} | \mathcal{H}_0 \right) = \mathbb{P} \left( \Re \{ \bm{y}^H \bm{\hat{\mu}} \} > \eta^{\prime} | \mathcal{H}_0 \right).
\end{equation}

The probability of detection is given by:
\begin{equation}
    P_{\text{D}} = \mathbb{P} \left( \Lambda^{\prime} > \eta^{\prime} | \mathcal{H}_1 \right) = \mathbb{P} \left( \Re \{ \bm{y}^H \bm{\hat{\mu}} \} > \eta^{\prime} | \mathcal{H}_1 \right).
\end{equation}

\textit{Remark 1:} The distribution of the GLRT statistic is:  Under hypothesis $\mathcal{H}_0$,  $ \Lambda^{\prime} \sim \mathcal{N}(0,\frac{\lVert \bm{\hat{\mu}} \lVert ^2}{2})$ and $ \Lambda^{\prime} \sim \mathcal{N}(\lVert \bm{\hat{\mu}} \lVert ^2,\frac{\lVert \bm{\hat{\mu}} \lVert ^2}{2})$ under hypothesis $\mathcal{H}_1$. This is proved in Appendix \ref{app:FI}. 

\subsection{Energy-based detection}

Another detection method we use is energy-based \cite{urkowitz1967energy}. Without estimating the DOA, the energy-based detection method saves computational effort. We formulate the test statistic for energy-based detection as:
$ V = \sum_{t=1}^{T} |y_t|^2.$

\textcolor{black}{
Under $\mathcal{H}_0$, since $y_t \sim \mathcal{CN}(0, \sigma^2)$, the real and imaginary components of $y_t$ each have variance $\sigma^2/2$. It follows that $\frac{2}{\sigma^2} V$ obeys a chi-square distribution with $2T$ degrees of freedom:$\frac{2}{\sigma^2} V \sim \chi^2(2T).$ Similarly, under $\mathcal{H}_1$, the statistic $V$ follows a non-central chi-square distribution:$\frac{2}{\sigma^2} V \sim \chi'^2\left(2T,\, P_s\right),$ where $P_s = \frac{2}{\sigma^2}\sum_{t=1}^{T} |\mu_t|^2$ is the non-centrality parameter representing the received signal energy. }
For a given threshold $\gamma$ we can derive the $P_{\text{FA}}$ and $P_{\text{D}}$ as:
\begin{equation}
\begin{aligned}
    P_{\text{FA}} &= \mathbb{P} \left(  V > \gamma | \mathcal{H}_0 \right),\\
    P_{\text{D}} & =  \mathbb{P} \left(  V > \gamma | \mathcal{H}_1 \right).
\end{aligned}
\end{equation}
The detection and false alarm probabilities can thus be obtained numerically via MATLAB functions (e.g., chi2cdf and ncx2cdf).

\subsection{Numerical results}

\begin{figure}[t]
  \centering
    \includegraphics[width=1.0 \linewidth]{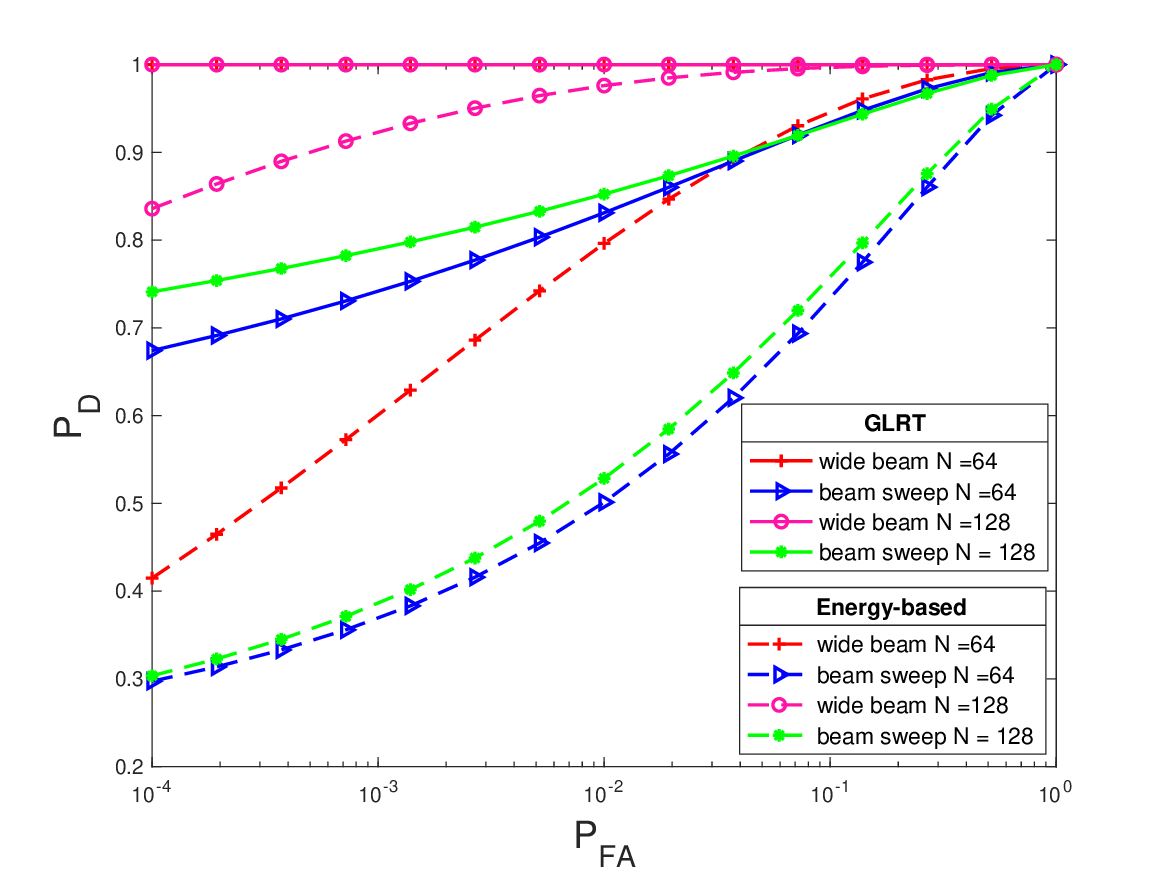}
  \caption{ GLRT \& Energy-based ROC curve with comparison between the beam sweeping method and the proposed method. Parameter: SNR = -15 dB.}
  \label{fig:detect}
    \vspace{-0.6cm}
\end{figure}

We use 5000 Monte Carlo simulations to generate random objects within $[-30^{\circ},30^{\circ}]$, and generate the average receiver operating characteristic (ROC) curve. We choose the number of antennas in RIS for $N =64,128$, and the number of time slots is set to $T = 7$. The simulation results are in Fig.~\ref{fig:detect}. 

We can see the different performance of the GLRT and energy-based detection methods. It is clear that
having an estimate of the unknown parameter significantly improves the detector performance. 
\textcolor{black}{However, this improved detection capability comes at the expense of computational complexity. Specifically, MUSIC requires the construction and eigenvalue decomposition of a $T \times T$ covariance matrix, which incurs $\mathcal{O}(T^3)$ operations, in addition to an angular spectrum search. In contrast, the energy detector only involves $T$ complex magnitude-squared operations, i.e., $\mathcal{O}(T)$ complexity. This highlights a fundamental trade-off between detector performance and computational cost, particularly relevant in real-time or power-constrained deployments.}

Figure~\ref{fig:detect} shows that the proposed wide beam method consistently outperforms the beam sweeping baseline, especially as the RIS size increases. \textcolor{black}{This is because the wide beam maintains uniform power over the entire ROI, ensuring reliable detection regardless of the target's direction. In contrast, beam sweeping relies on a limited number of narrow beams. With fixed angular sampling (e.g., \(10^\circ\) spacing for \(T=7\)), targets located outside the main lobes of the sweeping beams receive insufficient power.} 

\begin{figure}[t]
    \centering
    \includegraphics[width=1.0\linewidth]{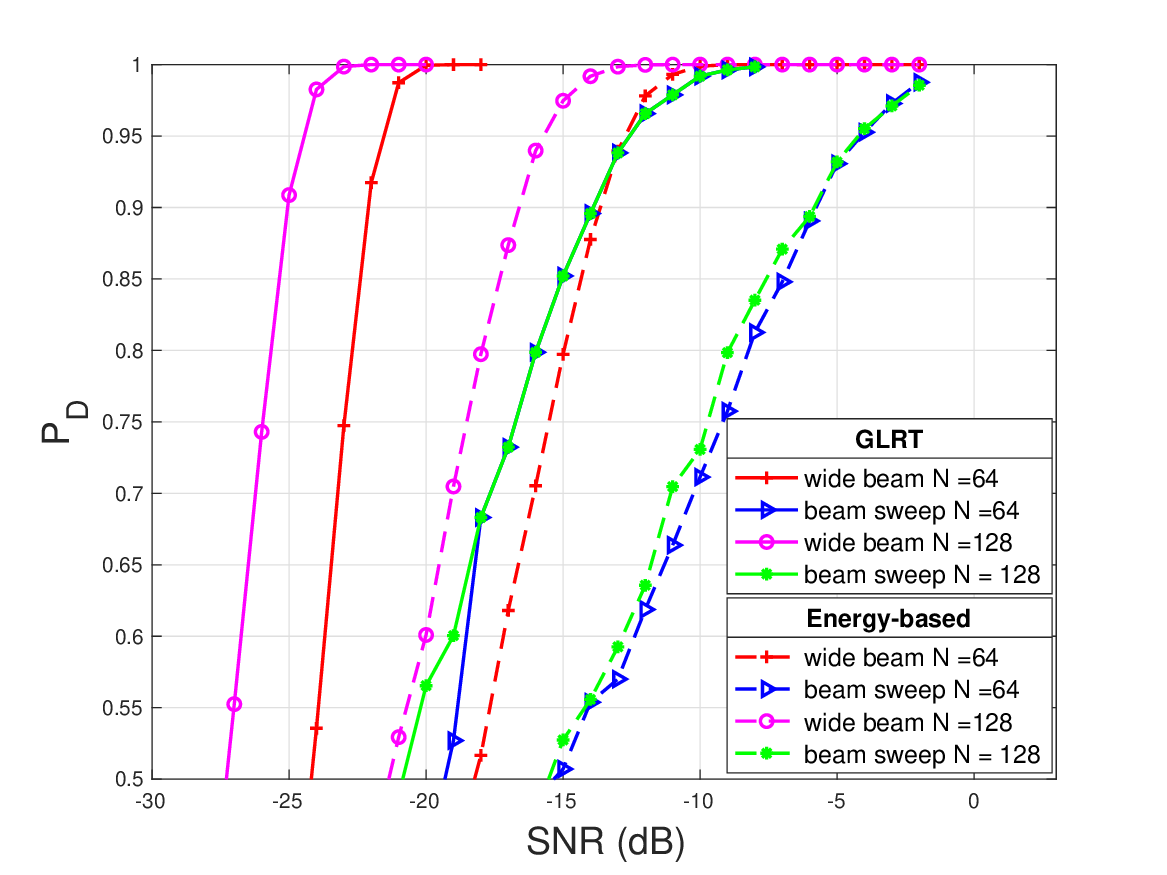}
    \caption{GLRT \& Energy-based detection performance with fixed $P_{\rm FA} = 0.01$ of beam sweeping and the proposed method. }
    \label{fig:fixpfa}
      \vspace{-0.6cm}
\end{figure}

Figure~\ref{fig:fixpfa} plots the detection probability $P_{\rm D}$ versus SNR at a fixed false-alarm probability of $P_{\rm FA} = 0.01$. As expected, the proposed wide beam method consistently outperforms the beam sweeping baseline. For instance, to achieve $P_{\rm D} = 0.9$ with a 64-element RIS, the wide beam requires roughly 8 dB less SNR than beam sweeping.
\textcolor{black}{Additionally, it is observed that the performance of the beam sweeping method doesn't improve with a larger number of elements $N$.  This is attributed to the narrower beams formed with a larger $N$, while the number of sweeping directions remains fixed at $T=7$. As a result, more of the angular region lies outside the main lobes, increasing the risk of missing the target. 
In contrast, the proposed wide beam method scales favorably with the RIS size. As shown in Fig.~\ref{fig:N}, increasing \(N\) boosts both the in-beam power and the target-to-side power ratio. This indicates that the proposed wide beam design remains robust under practical RIS size expansions.}

 \begin{figure*}[t]
  \centering
  \begin{subfigure}[b]{0.24\textwidth}
      \centering
      \includegraphics[width=\textwidth]{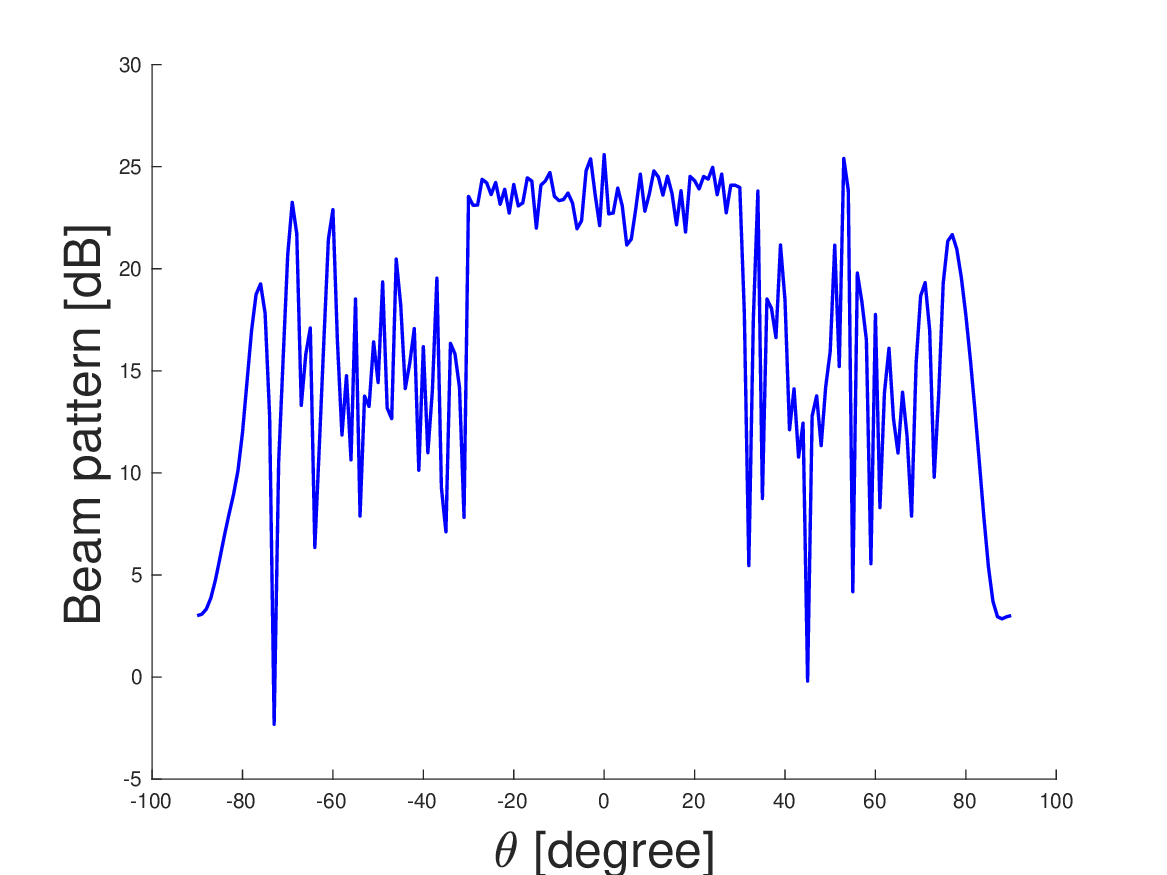}
      \caption{$\Delta \theta = 1^{\circ}$, $\Delta S = 1^{\circ}$ }
      \label{fig:theta1}
  \end{subfigure}
  \begin{subfigure}[b]{0.24\textwidth}
      \centering
      \includegraphics[width=\textwidth]{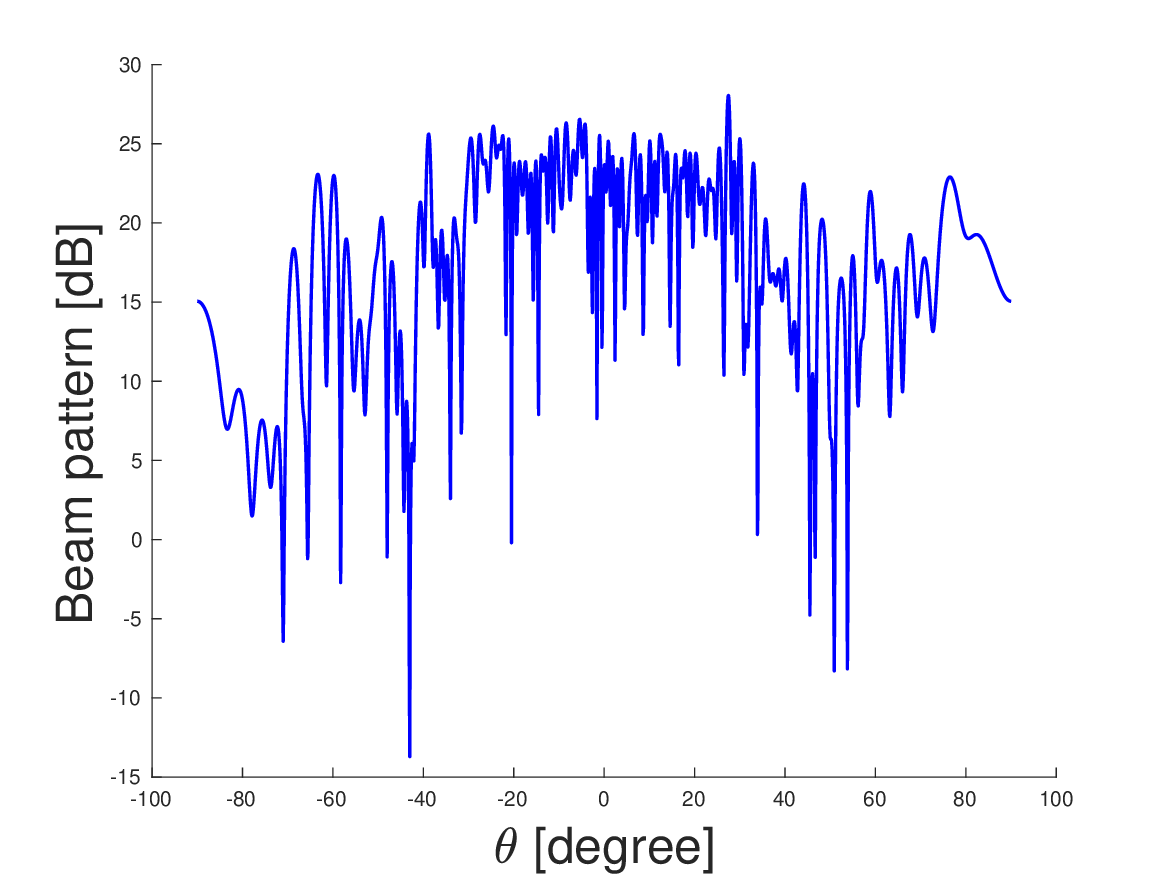}
      \caption{$\Delta \theta = 1^{\circ}$, $\Delta S = 0.1^{\circ}$}
      \label{fig:theta3}
  \end{subfigure}
  \begin{subfigure}[b]{0.24\textwidth}
      \centering
      \includegraphics[width=\textwidth]{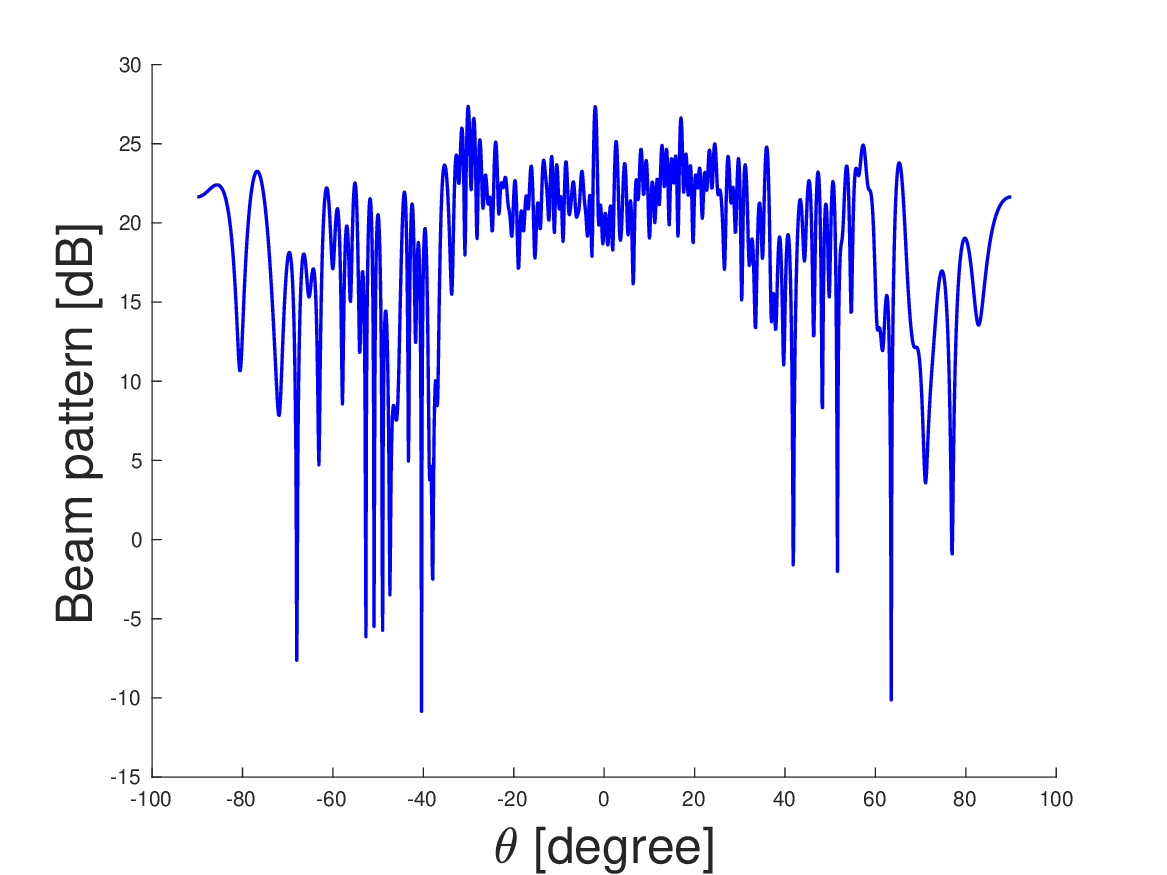}
      \caption{$\Delta \theta = 0.1^{\circ}$, $\Delta S = 0.1^{\circ}$}
      \label{fig:theta2}
  \end{subfigure}
  \begin{subfigure}[b]{0.24\textwidth}
      \centering
      \includegraphics[width=\textwidth]{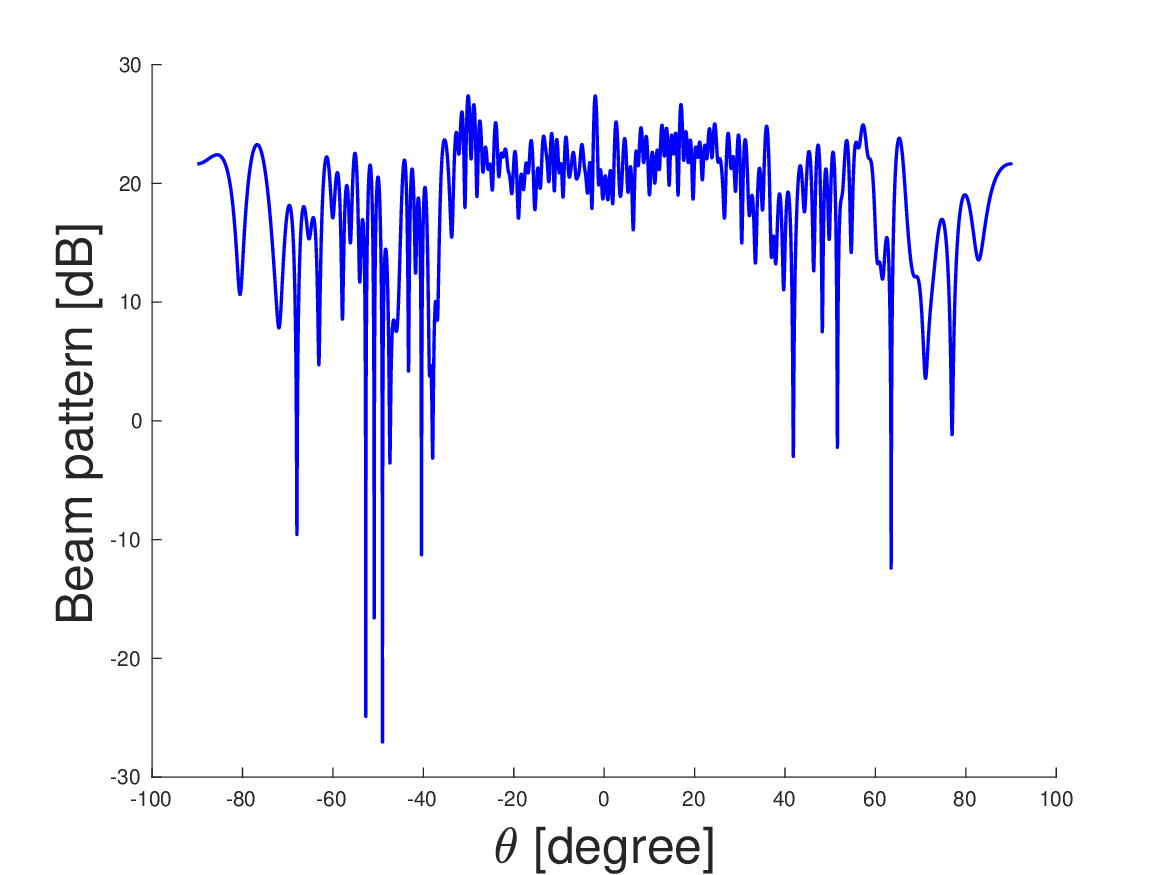}
      \caption{$\Delta \theta = 0.1^{\circ}$, $\Delta S = 0.01^{\circ}$}
      \label{fig:theta4}
  \end{subfigure}
  \caption{Beam pattern drawn under different discretization of $\theta$. Parameters: $N =128$, $L = 4$.}
  \label{fig:deltatheta}
    \vspace{-0.6cm}
\end{figure*}

\section{Conclusions}
\label{sec6}

This paper addresses the wide beam synthesis for RIS under discrete phase constraints when the target direction is unknown. We formulated the design as a max-min optimization problem and proposed a penalty-based MM algorithm to handle the non-convex discrete constraints while maintaining computational tractability.

Numerical results demonstrate significant improvements over beam sweeping: 8 dB SNR gain in AOA estimation and superior detection performance under both GLRT and energy-based criteria. The method provides power levels comparable to full amplitude-phase control systems with continuous amplitude and phase control while using only discrete phases. Moreover, the proposed method offers operational advantages through a wide angular coverage with only a single pilot, angular coverage thus reducing training overhead.
Future work may explore adaptive beam width design that dynamically adjusts the angular coverage based on previous observations and environmental feedback, enabling intelligent trade-offs between coverage and focusing gain.

 

\appendices

\section{Discretization Analysis}
 \label{appendix:discretization}

This appendix provides the analysis and experimental results supporting Remark~\ref{prop:discretization}.
To handle the infinite continuum of $\theta$ in problem (\ref{eq:opt01}), we discretize the angle with a step size $\Delta\theta$. The discretization should be fine enough to capture the beam’s characteristics and ensure that the synthesized beam pattern meets a desired flatness criterion (e.g., the power variation is no larger than -0.5 dB from the peak).  A convenient and conservative choice is to take $\Delta\theta$ no larger than the beamwidth corresponding to the target power level..
Figure~\ref{fig:deltatheta} illustrates the effect of different discretizations. For example, when using a coarse discretization of $\Delta\theta = 1^\circ$, the beam pattern appears uniform when plotted at the same resolution (Figs.\ref{fig:theta1}). However, when evaluated with a finer step size $\Delta S = 0.1^\circ$, significant variations become apparent (as shown in Fig.\ref{fig:theta3}). 

The underlying cause of this issue lies in the array factor $AF$, which is the impact that we can exert through beamforming on the array. For example, consider a Uniform Linear Array (ULA), whose array factor can be expressed as \cite{richards2005fundamentals}:
\begin{equation}
\label{eq:af}
    AF[\theta, \Delta\Phi] = \frac{\sin\left( N \left[ \frac{\pi d}{\lambda} \sin(\theta) - \frac{\Delta\Phi}{2} \right] \right)}{N \sin\left( \frac{\pi d}{\lambda} \sin(\theta) - \frac{\Delta\Phi}{2} \right)}
\end{equation} 
where $N$ is the number of antenna elements, $d $ is the inter-element spacing, $\Delta\Phi = \frac{2\pi d \sin\theta_0}{\lambda}$ is the phase shift between those adjacent elements and $\theta_0$ is the beam angle. The variable $\theta$ represents $AF$ changes over $\theta \in [-90^\circ, 90^\circ]$.

Let us assume the array elements is half-wavelength placing, and to simply the analysis, we assume the beam's boresight is $\theta_0 =0$, $AF$ in \eqref{eq:af} is simplified to:  
\begin{equation}
\label{eq:afq}
    AF_0[\theta] = \frac{\sin\left( N \frac{\pi }{2} \sin(\theta)  \right)}{N \sin\left( \frac{\pi }{2} \sin(\theta) \right)}.
\end{equation}
Following the definition of the half-power beamwidth (HPBW), which is then defined by finding the angle $\theta_\text{HPBW}$ such that its power shrinks to half of power at boresight $\theta_0$.
$\left|AF_0(\theta_\text{HPBW})\right| = \frac{1}{\sqrt{2}}.$
Similarly, we can also find the beamwidth related to, i.e., -1 dB power drop,
$
\left|AF_0\left(\theta_{-1\text{dB}}\right)\right| = 10^{-\frac{1}{20}}.
$
Table~\ref{table:beamwidth} lists the beamwidth values of a coherent ULA at several power drop levels.
Figures. \ref{fig:theta2} and \ref{fig:theta4} use a finer discretization of $\Delta\theta = 0.1^\circ$ during the optimization. Even when the beam pattern is plotted at a very fine step size of $ \Delta S = 0.01^\circ$ (Fig.\ref{fig:theta4}), the power remains relatively smooth, confirming that a sufficiently fine discretization is essential for producing a genuinely flat beam.
For the UPA case, the discretization steps must be chosen to be smaller than the corresponding beamwidths in both azimuth and elevation planes.
\begin{table}[h!]
\centering
\begin{tabular}{|c|c|c|c|}
\hline
\textbf{N } & \textbf{-3dB (HPBW) [$^\circ$]} & \textbf{-1dB [$^\circ$}] & \textbf{-0.5dB [$^\circ$]} \\ \hline
64 & 1.6 & 0.935 & 0.666 \\ \hline
100 & 1.0 & 0.6 & 0.4263 \\ \hline
128 & 0.8 & 0.4682 & 0.333 \\ \hline
256 & 0.4 & 0.2341 & 0.166 \\ \hline
\end{tabular}
\caption{Beam width values for a ULA  with different numbers of elements N and power levels.}
\label{table:beamwidth}
\vspace{-0.3cm}
\end{table}

\section{\textcolor{black}{Proof of Proposition~\ref{pp1}}}
\label{app:p1}
\begin{proof}
Let $\mathcal{S}\subset\mathbb C$ with $|s|=\rho$ for all $s\in\mathcal S$, and define
\[
\mathcal A=\{\bm x\in\mathbb C^N:\ x_i\in\mathcal S\},\qquad
\mathcal A'=\operatorname{conv}(\mathcal A).
\]
Every $\bm v\in\mathcal A$ satisfies $\|\bm v\|_2^2=N\rho^2$. Let $F(\bm x)=f(\bm x)+\lambda\|\bm x\|_2^2$ with $f$ convex and $\lambda\ge0$.

For any $\bm x\in\mathcal A'$, we can write
$\bm x=\sum_k \beta_k \bm v_k,$
with $\bm v_k\in\mathcal A$, $\beta_k\ge0$, $\sum_k\beta_k=1$. By convexity of $f$ and of $\|\cdot\|_2^2$ we obtain
\[
f(\bm x)\le \sum_k \beta_k f(\bm v_k),\qquad
\|\bm x\|_2^2\le \sum_k \beta_k \|\bm v_k\|_2^2=N\rho^2,
\]
hence
\[
F(\bm x)\le \sum_k \beta_k f(\bm v_k)+\lambda N\rho^2
          =\sum_k \beta_k F(\bm v_k)
          \le \max_{\bm v\in\mathcal A}F(\bm v).
\]
Therefore
\begin{equation}
\label{eq:ub}
\sup_{\bm x\in\mathcal A'}F(\bm x)\ \le\ \max_{\bm v\in\mathcal A}F(\bm v).
\end{equation}

On the other hand, since $\mathcal A\subseteq\mathcal A'$, we have 
\begin{equation}
\label{eq:lb}
\sup_{\bm x\in\mathcal A'}F(\bm x)\ \ge\ \max_{\bm v\in\mathcal A}F(\bm v).
\end{equation}

From \eqref{eq:ub}–\eqref{eq:lb}, the two quantities coincide, and the common value is attained because $\mathcal A$ is finite. Thus there exists $\bm v^\star\in\mathcal A$ such that
\[
\max_{\bm x\in\mathcal A'}F(\bm x)=\max_{\bm v\in\mathcal A}F(\bm v)=F(\bm v^\star).
\]

Now let $\bm x^\star\in\arg\max_{\bm x\in\mathcal A'}F(\bm x)$. Then
\[
F(\bm x^\star)=\max_{\bm x\in\mathcal A'}F(\bm x)
              =\max_{\bm v\in\mathcal A}F(\bm v),
\]
and the inequality \eqref{eq:ub} applied to $\bm x^\star$ shows that $\bm x^\star$ attains the upper bound given by the values of $F$ on $\mathcal A$. This implies that $\bm x^\star$ is also a maximizer at the same level as the points in $\mathcal A$, so
\[
\arg\max_{\bm x\in\mathcal A'}F(\bm x)\ \subseteq\ \arg\max_{\bm x\in\mathcal A}F(\bm x).
\]

Conversely, any maximizer over $\mathcal A$ is feasible for $\mathcal A'$ and reaches the same maximal value, hence
\[
\arg\max_{\bm x\in\mathcal A}F(\bm x)\ \subseteq\ \arg\max_{\bm x\in\mathcal A'}F(\bm x).
\]
Combining the two inclusions yields
\[
\arg\max_{\bm x\in\mathcal A'}F(\bm x)=\arg\max_{\bm x\in\mathcal A}F(\bm x).
\]
\end{proof}

\section{Proof of Proposition \ref{pp2}}
\label{app2}
\begin{proof}
    Given the MM algorithm’s update rule, at each iteration $k$, we solve the following optimization problem:
\begin{equation}
\label{eq:opt9}
\begin{aligned}
\max_{\bm{w} \in \mathcal{S}^{\prime}} \quad &  T \\
\text{s.t.} \quad &  T \leq g_\theta(\bm{w}| \bm{w^-}),\forall \theta \in [\theta_{min},\theta_{max}]. \\
\end{aligned}
\end{equation}
The update for the next iteration $\bm{w}^{(k+1)}$ is the solution of (\ref{eq:opt8}).
To prove the monotonic increase of $T$, consider:
\begin{equation}
\label{eq:monotonic}
\begin{aligned}
T(\bm{w}^{(k+1)}) & = \min_{\theta} f_\theta(\bm{w}^{(k+1)})
\geq \min_{\theta} g_\theta(\bm{w}^{(k+1)}| \bm{w}^{(k)}) \\
&\geq \min_{\theta} g_\theta(\bm{w}^{(k)}| \bm{w}^{(k)})
= \min_{\theta} f_\theta(\bm{w}^{(k)}) \\
&= T(\bm{w}^{(k)}).
\end{aligned}
\end{equation}

The first inequality follows from the fact that $g_\theta(\bm{w}^{(k+1)}| \bm{w}^{(k)})$ is a lower bound of $f_\theta(\bm{w}^{(k+1)})$ as given by equation (\ref{eq:lowbound}). The second inequality holds because $\bm{w}^{(k+1)}$ is the optimal solution of the optimization problem (\ref{eq:opt8}), which maximizes the value $T^{\prime}$, thereby increasing the minimum among the constraints. Thus, $T(\bm{w}^{(k+1)}) \geq T(\bm{w}^{(k)})$.
Therefore, the objective function $T$ is guaranteed to monotonically increase at each iteration of the MM algorithm.

Next, we want to prove that the sequence $ {T(\bm{w}^{(k)})}$  is bounded above. This can be shown by recognizing that $T(\bm{w})$ represents the objective function value which is bounded above by the maximum value of the original function  $f_\theta(\bm{w})$, given that $f_\theta(\bm{w})$ is a convex function over a convex hull, it must attain its maximum value within this set. Therefore, there exists some upper bound $T^{\rm UB}$ such that
$ T(\bm{w}) \leq T^{\rm UB}$.

\textcolor{black}{In addition to the monotonicity and boundedness proven above, the convergence of the sequence $\{\bm{w}^{(k)}\}$ follows from standard MM theory. As established in \cite[Theorem 1]{sunMajorizationMinimizationAlgorithmsSignal2017}, if the surrogate function is continuously differentiable, globally bounds the objective (as in our construction), and is tight at the current iterate, then the MM iterates converge to a first-order stationary point. These conditions are satisfied by our surrogate in Eq.~\eqref{eq:lowbound}, ensuring that the proposed MM-based algorithm converges to a stationary point of the relaxed problem.}

\end{proof}

\section{The distribution of the test statistic of GLRT}
\label{app:FI}
\subsubsection{$P_{\text{FA}}$ under Hypothesise $\mathcal{H}_0$}
To start with, we normalize the noise power into $\sigma^2 =1$. Following the signal model under $\mathcal{H}_0$, we have $y_i \sim \mathcal{CN}(0,1)$, for each independent observation.
Let us rewrite $y_i$ into its real and imaginary part, we have $y_i = x_i + j z_i$ and $x_i \sim \mathcal{N}(0,1/2)$, $z_i \sim \mathcal{N}(0,1/2)$. Similarly, we rewrite the estimation  $\bm{\hat{\mu}} = [\mu_1, \cdots \mu_i,\cdots, \mu_T ]^T$ and  each $\mu_i = a_i +j b_i$.

So, the test statistic can be expanded as:
\begin{equation*}
\Lambda^{\prime} = \Re\{ \bm{y}^H \bm{\hat{\mu}} \} = \sum_{i=1}^{T} a_i x_i + b_i z_i
\end{equation*}

The mean value is zero $ \mathbb{E}[\Lambda^{\prime} ] = 0$ since  $x_i$  and $ z_i$  are both zero-mean normally distributed, and the weighted sum maintains a zero mean, and the variance is $
\text{Var}\left( \Lambda^{\prime} \right) = \frac{1}{2} \sum_{i=1}^{T} \left( a_i^2 + b_i^2 \right) = \frac{1}{2} \lVert \bm{\hat{\mu}} \lVert^2$. So under $\mathcal{H}_0$, the test statistic $ \Lambda^{\prime} \sim \mathcal{N}(0,\frac{\lVert \bm{\hat{\mu}} \lVert ^2}{2})$.

The corresponding value of $P_{\text{FA}}$ is given by the following:
\begin{equation}
   P_{\text{FA}} = \mathbb{P} \left( \Lambda^{\prime} > \eta^{\prime} | \mathcal{H}_0 \right) = Q(\frac{\eta^{\prime}}{\sqrt{\frac{\lVert \bm{\hat{\mu}} \lVert ^2}{2}}}). 
\end{equation}
To archive a certain $P_{\text{FA}}$ the decision threshold is:
\begin{equation}
     \eta^{\prime} = \sqrt{\frac{\lVert \bm{\hat{\mu}} \lVert ^2}{2}} Q^{-1}(P_{\text{FA}})
\end{equation}
 where $Q(x) = \int_{x}^{+\infty} \frac{1}{\sqrt{2\pi}} e^{-\frac{1}{2}t^2} dt$, and $ Q^{-1}(x)$ is the inverse integral function.

\subsubsection{$P_{\text{D}}$ under hypothesise $\mathcal{H}_1$}

As $y_i \sim \mathcal{CN}(\hat{\mu}_i,1)$, its real and imaginary part follows $x_i \sim \mathcal{N}(a_i,1/2)$, $z_i \sim \mathcal{N}(b_i,1/2)$. 
The corresponding mean and variance are:
\begin{equation*}
\begin{aligned}
    \mathbb{E}[\Lambda^{\prime}] &= \sum_{i=1}^{N} |\hat{\mu}_i|^2 = \lVert \bm{\hat{\mu}} \lVert ^2 \\
    \text{Var}\left( \Lambda^{\prime} \right) &= \frac{1}{2} \sum_{i=1}^{T} \left( a_i^2 + b_i^2 \right) = \frac{1}{2} \lVert \bm{\hat{\mu}} \lVert^2
\end{aligned}  
\end{equation*}
So we have $ \Lambda^{\prime} \sim \mathcal{N}(\lVert \bm{\hat{\mu}} \lVert ^2,\frac{\lVert \bm{\hat{\mu}} \lVert ^2}{2})$, and $P_{\text{D}}$ is given as:
\begin{equation}
    P_{\text{D}} = \mathbb{P} \left( \Lambda^{\prime} > \eta^{\prime} | \mathcal{H}_1 \right) = Q(\frac{\eta^{\prime}- \lVert \bm{\hat{\mu}} \lVert ^2}{\sqrt{\frac{\lVert \bm{\hat{\mu}} \lVert ^2}{2}}}). 
\end{equation}
\textcolor{black}{
Combining the results under $\mathcal{H}_0$ and $\mathcal{H}_1$, we can eliminate the threshold $\eta^{\prime}$ and express the receiver operating characteristic (ROC) in closed form as:}
\begin{equation}
    P_{\text{D}} =  Q(Q^{-1}(P_{\text{FA}}) - \lVert \bm{\hat{\mu}} \lVert ^2). 
\end{equation}

\bibliographystyle{ieeetr}
\bibliography{strings}

\end{document}